\documentclass[sort&compress,english,preview,5p]{elsarticle}
\usepackage{graphicx}
\usepackage{dcolumn}
\usepackage{bm}
\usepackage{amsmath}
\usepackage{comment}
\usepackage{epstopdf} 
\usepackage{xr-hyper}
\usepackage{lineno,hyperref}
\usepackage{color} 
\usepackage{soul} 

\makeatletter

\usepackage{amssymb}

\usepackage{lineno,hyperref}
\modulolinenumbers[5]

\journal{Ultramicroscopy}

\renewcommand{\vec}[1]{\mathbf{#1}}

\begin{document}

\begin{frontmatter}



\title{Investigating the convergence properties of iterative ptychography for atomic-resolution low-dose imaging}

\author[emat,nanolab]{Tamazouzt Chennit\corref{corresp}}
\author[emat,nanolab]{Songge Li}
\author[emat,nanolab]{Hoelen L. Lalandec Robert}
\author[emat,nanolab]{Christoph Hofer}
\author[emat,nanolab]{Nadine J. Schrenker}
\author[iit]{Liberato Manna}
\author[emat,nanolab]{Sara Bals}
\author[emat,nanolab]{Timothy J. Pennycook}
\author[emat,nanolab]{Jo Verbeeck}

\cortext[corresp]{Corresponding author: tamazouzt.chennit@uantwerpen.be}
\address[emat]{EMAT, University of Antwerp, Groenenborgerlaan 171, 2020 Antwerp, Belgium}
\address[nanolab]{NANOlight Center of Excellence, University of Antwerp, Groenenborgerlaan 171, 2020 Antwerp, Belgium}
\address[iit]{Department of Nanochemistry, Istituto Italiano di Tecnologia (IIT), 16163 Genova, Italy}

\begin{abstract}
    This study investigates the convergence properties of a collection of iterative electron ptychography methods, under low electron doses ($<$ 10$^3$ $e^-/\text{\AA}^2$) and gives particular attention to the impact of the user-defined update strengths. We demonstrate that carefully chosen values for this parameter, ideally smaller than those conventionally met in the literature, are essential for achieving accurate reconstructions of the projected electrostatic potential. Using a 4D dataset of a thin hybrid organic-inorganic formamidinium lead bromide (FAPbBr$_{3}$) sample, we show that convergence is in practice achievable only when the update strengths for both the object and probe are relatively small compared to what is found in literature. Additionally we demonstrate that under low electron doses, the reconstructions initial error increases when the update strength coefficients are reduced below a certain threshold emphasizing the existence of critical values beyond which the algorithms are trapped in local minima. These findings highlight the need for carefully optimized reconstruction parameters in iterative ptychography, especially when working with low electron doses, ensuring both effective convergence and correctness of the result. 
\end{abstract}


\end{frontmatter}



\section{Introduction}
    
    Atomic resolution imaging of crystalline materials, typically requiring high electron doses \cite{egerton_dose_2021}, has become a routine procedure in transmission electron microscopy (TEM) \cite{erni_atomic-resolution_2009}, in particular, thanks to the widespread use of aberration correction \cite{batson_sub-angstrom_2002}. However, numerous specimens of interest are susceptible to beam damage, which makes high-resolution challenging \cite{chen_imaging_2020}. As described e.g. in references \cite{egerton_mechanisms_2012, egerton_control_2013, egerton_radiation_2019}, two damage mechanisms are prevalent in TEM. Those are radiolysis and knock-on displacement. Radiolysis primarily affects semiconductors and insulators, and results from the inelastic interaction between the electron beam and core electrons in the specimen, i.e. leading to ionization. Knock-on effects are typically more dominant in metals and result from elastic interactions, involving significant energy transfers to atoms and their displacement. This damage mechanism is particularly common in two-dimensional materials, where it often manifests as hole formation \cite{susi_quantifying_2019, meyer_accurate_2012}.
    
    Inorganic perovskites such as CsPbI$_{3}$ \cite{scheid_electron_2023, satta_formation_2021}, hybrid organic-inorganic perovskites such as methylammonium lead iodide (MAPBI$_{3}$) \cite{dhivyaprasath_degradation_2023} or formamidinium lead bromide (FAPbBr$_{3}$) \cite{schrenker_investigation_2024}, zeolites \cite{ugurlu_radiolysis_2011, liu_direct_2020, ooe_direct_2023} and metal-organic frameworks (MOF) \cite{furukawa_chemistry_2013, yang_catalysis_2019, bavykina_metalorganic_2020, kavak_high-resolution_2025} are prominent examples of beam-sensitive materials in the materials science domain. Moreover, dose-sensitivity is of significant importance in biological sciences \cite{glaeser_limitations_1971, henderson_three-dimensional_1975}, especially when imaging proteins \cite{kucukoglu_low-dose_2024} and viruses \cite{zhou_low-dose_2020}.
    
    Conventional TEM (CTEM) is widely used for low-dose imaging; however, methods based on scanning TEM (STEM), such as the integrated center of mass (iCoM) \cite{lazic_phase_2016, lazic_single-particle_2022, li_direct_2019, liu_direct_2020} and ptychography, have been used and have shown promising capacities as well \cite{yang_efficient_2015-1, oleary_phase_2020, oleary_contrast_2021, scheid_electron_2023, hao_atomic-scale_2023, dong_atomic-level_2023, zhou_low-dose_2020, robert_benchmarking_2025, li_atomically_2025, lozano_low-dose_2018,D'Alfonso_deterministic, D'Alfonso_dose_dependent}.
    
    Ptychography, belonging to the family of coherent diffractive imaging (CDI) techniques \cite{fienup_reconstruction_1978, fienup_phase_1982, fienup_reconstruction_1987, miao_phase_1998, miao_extending_1999}, was initially introduced by Hoppe et al. \cite{hoppe_beugung_1969, hoppe_beugung_1969-1, hoppe_beugung_1969-2} in 1969. Recently, there has been renewed interest in ptychographic methods that employ fast electrons, particularly with the advent of direct electron detectors (DED) \cite{mcmullan_electron_2007,llopart_timepix_2007, ballabriga_medipix3_2011, mir_characterisation_2017, ryll_pnccd-based_2016, philipp_very-high_2022, zambon_enhanced_2023, ercius_4d_2024}. These detectors, having frame rates of typically 10$^3$ to 10$^4$ per second, permit the acquisition of four-dimensional STEM (4D-STEM) datasets \cite{yang_4d_2015} with reasonable recording times, as required in ptychography, though with stability and scan size restrictions. Those last remaining limits were largely removed in the last few years, following the introduction of event-driven detectors \cite{poikela_timepix3_2014, llopart_timepix4_2022} enabling sub-microsecond dwell times in STEM \cite{jannis_event_2022, auad_time_2024, denisov_characterization_2023} and facilitating low-dose acquisitions. With those experimental improvements, ptychography was thus enabled as a powerful phase retrieval technique permitting atomic resolution with high dose efficiency.
    
    Ptychography includes analytical approaches such as the Wigner distribution deconvolution (WDD) \cite{rodenburg_theory_1992, bates_sub-angstrom_1989, mccallum_two-dimensional_1992} and single-sideband (SSB) \cite{rodenburg_experimental_1993, pennycook_efficient_2015, yang_4d_2015} methods, originating in the 1990s. An iterative variant was later introduced in the form of the ptychographic iterative engine (PIE) \cite{faulkner_movable_2004, rodenburg_phase_2004} which was furthermore extended to simultaneously refine both the object and the illumination, resulting in the extended version of PIE known as ePIE \cite{maiden_improved_2009, maiden_further_2017}. Other methods, such as error reduction (ER) \cite{fienup_phase_1982, thibault_probe_2009, thibault_high-resolution_2008} and weighted average sequential projection (WASP) \cite{maiden_wasp_2024}, have also been proven to be efficient. Iterative methods involve repeatedly updating both the specimen and the probe functions, with each update weighted by a user-defined update coefficient or strength.
    
    To date, only a few studies have explored the impact of these update coefficients across a range of iterative algorithms. One notable example is \cite{maiden_further_2017}, which compares various ePIE-like algorithms. However, that study was not specifically aimed at achieving atomic resolution, nor did it address the challenges associated with low electron doses. In this manuscript, we fill this gap by investigating the convergence behavior of these iterative methods under low electron dose conditions, with a particular focus on how the choice of update strengths influences their performance. Using both simulated and experimental 4D datasets from a FAPbBr$_{3}$ nanocrystal, we show that the convergence dynamics of ePIE-like algorithms are influenced not only by acquisition parameters such as electron dose but also by external factors like the update coefficients. Our findings indicate that the performance generally degrades beyond a certain threshold of update coefficients, suggesting that these iterative algorithms are sensitive to such external parameters. Moreover, we find that the regularized version of ePIE (rPIE) demonstrates greater robustness and more stable convergence across a range of update strengths compared to the standard ePIE algorithm.
    
    The first section of this manuscript explores the fundamental principles of both analytical and iterative approaches to ptychography, while the second section provides a comparative evaluation of these techniques against the direct SSB solution and finally, the core of the manuscript is devoted to examining how the user-defined update coefficients affect both the quality of the reconstructions and the convergence behavior of the algorithms, especially under low electron dose conditions.

\section{Fundamental aspects of electron ptychography}
    
    \subsection{\textit{The phase problem}}
        
        \begin{figure}
            \centering
            \includegraphics[width=0.5\textwidth]{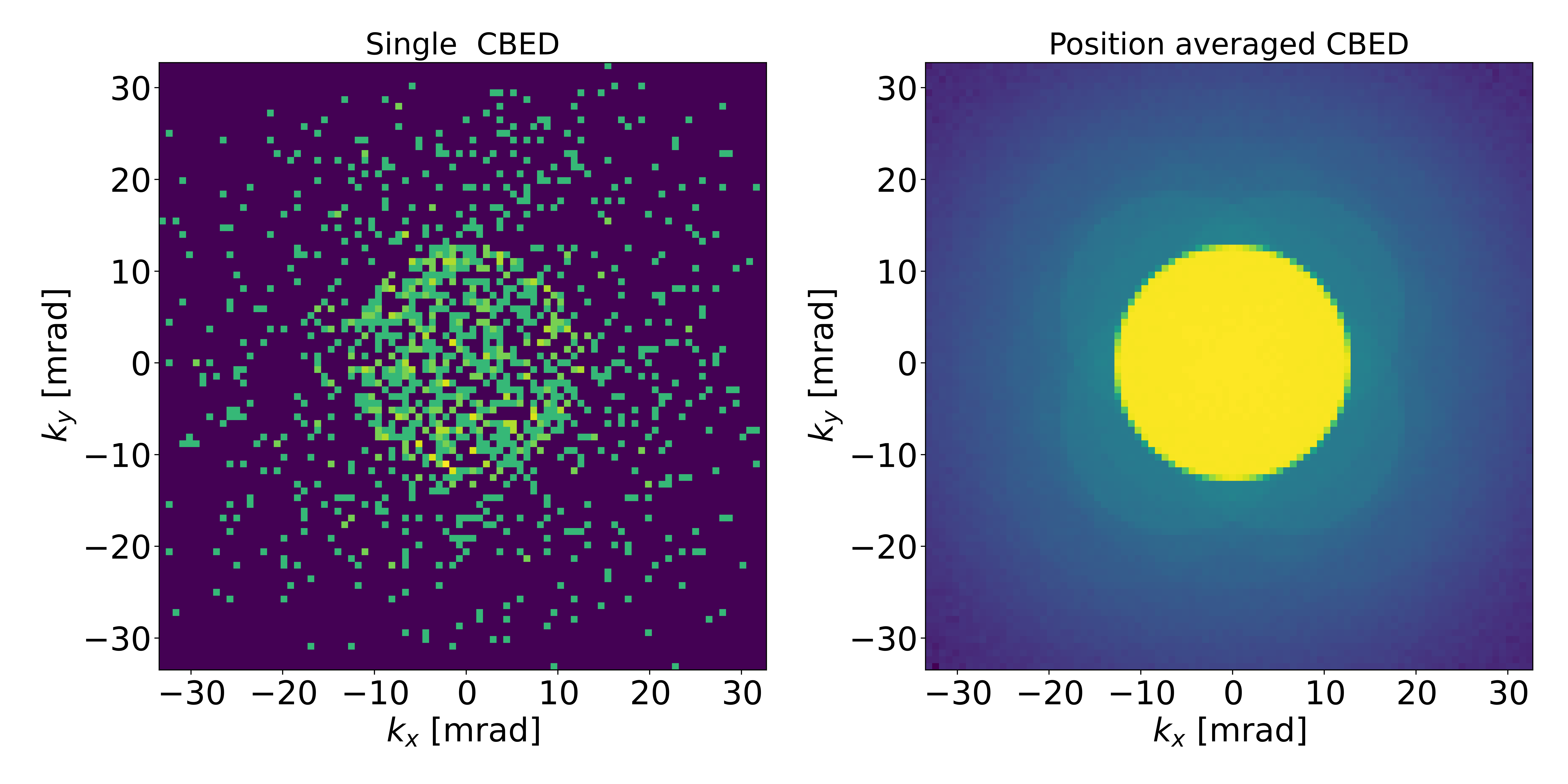}
            \caption{Position-averaged CBED (to the left) and single CBED (to the right) obtained from a simulated 4D dataset of 2.5 nm thick FAPbBr$_{3}$ using a maximum collection angle of 32 mrad, a convergence angle of 13 mrad, an acceleration voltage of 200 kV. The probe is aberration-free.}
            \label{fig-1}
        \end{figure}
        
        In a conventional 4D-STEM experiment, a well-focused electron probe is scanned over the sample and a convergent electron beam diffraction (CBED) pattern is recorded in the far-field by a DED. Figure \ref{fig-1} represents a simulated single CBED pattern and its scan position-averaged counterpart, obtained with a maximum collection angle of 32 mrad using a low electron count dataset showing Poisson counting noise.
        
        The recorded intensity in the far-field is given by:
        \begin{align}
            I_{\text{CBED}}(\vec{K}_r, \vec{R}_0) &= \left| \tilde{\psi}(\vec{K}_r, \vec{R}_0) \right|^2 \label{icbed}\\
            &= \left| \mathcal{F} \left[ \psi(\vec{r}, \vec{R}_0) \right] \right|^2 \nonumber \\
            &= \left| \mathcal{F} \left[ P(\vec{r}-\vec{R}_0) \, O(\vec{r}) \right] \right|^2 \nonumber
        \end{align}
        
        In such an experiment, the measurable intensity\\ \( I_{\text{CBED}}(\vec{K}_r, \vec{R}_0) \) given by equation \ref{icbed}corresponds to the squared modulus of the Fourier transform of the real-space exit wave. For a relatively thin specimen, where the phase object approximation (POA) holds, as utilized e.g. in ref. \cite{cowley_electron_1972, rodenburg_theory_1992, rodenburg_phase_2004, maiden_improved_2009}, this exit wave can be expressed as the product of the object function \( O(\vec{r}) \) with the scanned probe \( P(\vec{r} - \vec{R_0}) \). Here \( \vec{R_0} \) denotes the relative shift between the probe's position and the object and r denotes the coordinate vector in the real space. This can be formally written as:
        \begin{align}
            \psi(\vec{r}, \vec{R}_0) &= P(\vec{r}-\vec{R}_0) \, O(\vec{r})
            \label{psifunction}
        \end{align}
        
        Here, \(\vec{K}_r\) denotes the reciprocal space coordinate vector, \( \psi(\vec{r}, \vec{R}_0) \) and \( \tilde{\psi}(\vec{K}_r, \vec{R}_0) \) are the exit wave at the specimen plane and its far-field projection, respectively. To remove the non-physical amplitude-scaling ambiguity between object and probe in iterative reconstruction, we normalize the probe power at each iteration so that its total intensity remains fixed. This ensures that amplitude is not arbitrarily traded between object and probe while leaving the exit wave unchanged. The global phase of the object remains arbitrary, as is standard in phase retrieval.
        
        As the recording consists of Poisson-distributed electron counts \cite{luczka_master_1991,seki_theoretical_2018} determined by the squared modulus of $\tilde{\psi}(\vec{K}_r, \vec{R}_0)$, the phase distribution of the far-field wave is not directly accessible. As a consequence, the spatially dependent, interaction-induced, phase shift experienced locally by the electron probe cannot be recovered from a single CBED pattern, except in limited conditions \cite{miao_phase_1998, weierstall_image_2002, martin_practical_2011,{morgan_fast_2013}}. This is often referred to as the \textit{``phase problem''} \cite{drenth_problem_1975} in optics and electron microscopy and presents a significant challenge, particularly when imaging at low electron doses, as the recorded intensities may be insufficient for unambiguous interpretation using conventional STEM methods.
        
        Continuing, the specimen transmission function \( O(\vec{r}) \) is given by:
        \begin{equation}
            O(\vec{r}) = M(\vec{r}) \, \exp \big( -i \, \theta(\vec{r}) \big)
            \label{object}
        \end{equation}
        Here, \( \theta(\vec{r}) \) represents the phase shift induced by the interaction with the specimen and \( M(\vec{r}) \) is an amplitude modulation term. Under the POA, it is postulated that the specimen induces changes to the incident electron wave solely in phase. Consequently, \( M(\vec{r}) \) can be assigned a constant value of 1, although in practice a lack of perfect coherence or the POA not being perfectly fulfilled will lead to a non-unitary amplitude in the result. The integration of the phase shift along the specimen's thickness yields a relationship between \( \theta(\vec{r}) \) and the projected electrostatic potential of the specimen \( V_t(\vec{r}) \) given by:
        \begin{equation}
            O(\vec{r}) = \exp \left( i \sigma V_{t}(\vec{r}) \right)
            \label{object-phase}
        \end{equation}
        The quantity $\sigma$ then represents the interaction coefficient.
        
        When the imaged object is sufficiently light in addition to being thin, such that the range of values covered by $\sigma V_{t}(\vec{r})$ is negligible compared to 1, the exponent in the object transmission function can be Taylor-expanded to the first-order. Consequently, $O(\vec{r})$ becomes linearly dependent on the projected potential. This approximation is referred to as the weak phase object approximation (WPOA).
    
    \subsection{\textit{Single sideband (SSB)}}
        
        The extraction of the phase of the object’s transmission function in the framework of the single sideband (SSB) algorithm, as initially described by Rodenburg et al. \cite{rodenburg_experimental_1993} and later tested for atomic resolution imaging \cite{pennycook_efficient_2015, oleary_phase_2020, songge_li_2025}, involves the calculation of the Fourier transform of the recorded intensity \( \big| \tilde{\psi}(\vec{K}_r, \vec{R}_0) \big|^2 \) with respect to the probe position \( \vec{R}_0 \). This can be written as:
        \begin{align}
            G(\vec{K}_r, \vec{Q}_0) &= \int \big| \tilde{\psi}(\vec{K}_r, \vec{R}_0) \big|^2 \, \exp \big( -2 i \pi \vec{R}_0 \vec{Q}_0 \big) \, d\vec{R}_0
            \label{gfunction1}
        \end{align}
        Under the WPOA, \( G(\vec{K}_r, \vec{Q}_0) \) is equal to:
        \begin{align}
            G(\vec{K}_r, \vec{Q}_0) &= \big| A(\vec{K}_r) \big|^2 \, \delta(\vec{Q}_0) \nonumber \\
            &+ A(\vec{K}_r) \, A^*(\vec{K}_r+\vec{Q}_0) \, i\tilde{\theta}(\vec{Q}_0) \nonumber \\
            &- A^*(\vec{K}_r) \, A(\vec{K}_r-\vec{Q}_0) \, i\tilde{\theta}(\vec{Q}_0)
            \label{gfunction2}
        \end{align}
        Here, \( A(\vec{K}_r) \) denotes the aperture function, which comprises both an amplitude and an aberration function \( \chi(\vec{K}_r) \) and \(\tilde{\theta}\) represents the Fourier transform of the phase shift.
        \begin{align}
            A(\vec{K}_r) &= \Omega \, \exp \left( i \, \chi(\vec{K}_r) \right) \\
            \Omega &= 1 \quad \text{inside the aperture} \nonumber  \\
            \Omega &= 0 \quad \text{outside the aperture} \nonumber
            \label{afunction}
        \end{align}
        
        In the triple overlap region where the $0^{th}$ order scattered disk and the first-order scattered disks overlap, the two last terms cancel out. As a result, this area cannot contribute to the retrieval of the spatially dependent phase shift. It is thus only achievable by exploitation of the double overlap region, between the $0^{th}$ and first-order scattered disks. In most cases, this is done by direct summation \cite{pennycook_efficient_2015}, though a deconvolutive solution exists as well \cite{yang_enhanced_2016,robert_benchmarking_2025}.
    
    \subsection{\textit{Iterative methods}}
        
        Iterative ptychography constitutes an optimization problem aiming to reconstruct accurate phase images of a specimen. It is conventionally initialized with an object possessing a constant amplitude and phase, respectively equal to 1 and 0, and with a non-aberrated probe.Initializing the object with unit amplitude and zero phase is a standard practice in iterative phase retrieval algorithms, as it provides a neutral starting point and ensures that the reconstruction is not biased by prior assumptions about the object's structure\cite{hawkes_ptychography_2019, maiden_improved_2009, maiden_superresolution_2011}.

        The original PIE algorithm \cite{rodenburg_phase_2004, faulkner_movable_2004} can generally be considered as originating from the Gerchberg-Saxton algorithm \cite{gerchberg_practical_1972, gerchberg_super-resolution_1974}, where the phase information is retrieved via the replacement of the calculated amplitude of the exit wave, typically in two distinct optical planes or more, by the experimental amplitude and/or a support constraint. In PIE, this concept is extended to the situation where a single optical plane is used for the acquisition of intensity, typically the far-field, and where illumination conditions are varied, specifically by shifting the probe. An update function is then defined for each probe position, capturing the difference between the original exit wave in real space and the one obtained after replacing the amplitude in reciprocal space. A multiplying factor, sometimes referred to as a regularizer \cite{maiden_further_2017} and proportional to the probe intensity, is also inserted to limit the updated area in the specimen plane. Iteratively adding this function to the reconstructed object, while treating scan positions sequentially and weighting it with a user-defined update coefficient, permits the retrieval of an object that satisfies equation \ref{pie} \cite{melnyk_connections_2022}. The extended version of PIE \cite{maiden_improved_2009} aims to find both the object and the probe via the same process. The error function is calculated at each inner iteration, which we define as the computation performed for a single probe position, and convergence is assessed by monitoring the evolution of the error metric over successive iterations. Here, an iteration refers to the computation of update functions across the entire batch of probe positions, starting from an initial randomly selected probe position $j=1$ and continuing until all positions in the batch $j=J$ have been processed.
        
        \begin{equation}
            \big| \tilde{\psi}(\vec{K}_r, \vec{R_{0}}) \big| = \sqrt{I_{\text{CBED}}(\vec{K}_r, \vec{R}_0)}
            \label{pie}
        \end{equation}
        
        Many iterative algorithms exist today \cite{rodenburg_phase_2004, guizar-sicairos_phase_2008, maiden_improved_2009, maiden_further_2017, odstrcil_iterative_2018, maiden_wasp_2024, thibault_maximum-likelihood_2012}, with one essential distinction among them being the method of update execution. In the original PIE and ePIE algorithms, the exit wave update is calculated for each probe position and applied individually, following a random order within the complete set of probe positions. This methodology is referred to as a sequential projection (SP) algorithm \cite{maiden_wasp_2024}. An alternative algorithmic approach, such as ER, involves computing the complete updates of the exit waves prior to their incorporation into the probe and object functions. This is termed as a batch processing (BP) type of algorithm. A hybrid processing (HP) variant exists, known as the WASP \cite{maiden_wasp_2024}, which integrates the update calculation method of ER algorithms with the input methodology of ePIE-like algorithms.
        
        Most SP algorithms, such as ePIE, follow a workflow that involves propagating the exit wave, given in equation \ref{psifunction}, to the far field by calculating its Fourier transform. The calculated modulus is in turn replaced with the experimental measurement. The corrected wave is then propagated back to real space by calculating its inverse Fourier transform. The final step involves updating the object and probe functions using the following equations:
        \begin{align}
            O_{j+1}(\vec{r}) &= O_{j}(\vec{r}) \; + \; \alpha \; \frac{P_{j}^{*}(\vec{r})}{\big|P_{j}(\vec{r}) \big|_{max}^{2}} \; (\psi_{j}^{'}(\vec{r}) \; - \; \psi_{j}(\vec{r})) \\
            P_{j+1}(\vec{r}) &= P_{j}(\vec{r}) \; + \; \beta \; \frac{O_{j}^{*}(\vec{r})}{\big|O_{j}(\vec{r}) \big|_{max}^{2}} \; (\psi_{j}^{'}(\vec{r}) \; - \; \psi_{j}(\vec{r}))
            \label{pie-update}
        \end{align}
        $O_{j}(\vec{r})$ represents the current estimate of the object function at probe position \( j \), while \( O_{j+1}(\vec{r}) \) represents the updated object function, that will be used as the estimate for the next calculation. This process continues until all probe positions within the batch of size (J) are taken into account. $\psi_{j}(\vec{r})$ and $\psi_{j}^{'}(\vec{r})$ represent the estimated and updated exit waves, respectively, in real space.
        
        The update strengths $\alpha$ and $\beta$ regulate the extent to which the updated probe and object functions deviate from their previous states and typically, multiple iterations are necessary for full convergence of the algorithm. The order in which each position is treated is semi-random, as each one them is visited only once. The required number of updates performed is in practice determined by the minimization of an error function in case of a simulated 4D dataset in which the ground truth is already known. In this context, we calculate the sum squared error (SSE).
        
        \begin{align}
            SSE &= \frac{1}{J} \sum_{j=1}^{J} \left( \big| \psi_{j}^{'}(\vec{K}_r, \vec{R}_0) \big|^{2} - \big| \psi_{j}(\vec{K}_r, \vec{R}_0) \big|^{2} \right)^{2} 
            \label{SSE}
        \end{align}
        
        The exploration of more efficient regularizers than those used in ePIE led to the development of the regularized ptychographic iterative engine (rPIE). The object and probe updates for rPIE are given by equation \ref{rpie-update-object} and \ref{rpie-update-probe} respectively.
        
        \begin{align}
            O_{j+1}(\vec{r}) &= O_{j}(\vec{r}) \; + \; \frac{P_{j}^{*}(\vec{r}) \; (\psi_{j}^{'}(\vec{r}) \; - \; \psi_{j}(\vec{r}))}{(\alpha \; - \; 1) \; \big|P_{j}(\vec{r}) \big|^{2} \; + \; \alpha \; \big|P_{j}(\vec{r}) \big|_{max}^{2}} \;
            \label{rpie-update-object}
        \end{align}
        \begin{align}
            P_{j+1}(\vec{r}) &= P_{j}(\vec{r}) \; + \; \frac{O_{j}^{*}(\vec{r}) \; (\psi_{j}^{'}(\vec{r}) \; - \; \psi_{j}(\vec{r}))}{(\beta \; - \; 1) \; \big|O_{j}(\vec{r}) \big|^{2} \; + \; \beta \; \big|O_{j}(\vec{r}) \big|_{max}^{2}} \;
            \label{rpie-update-probe}
        \end{align}

        \begin{figure}[h!]
            \centering
            \includegraphics[width=0.48\textwidth]{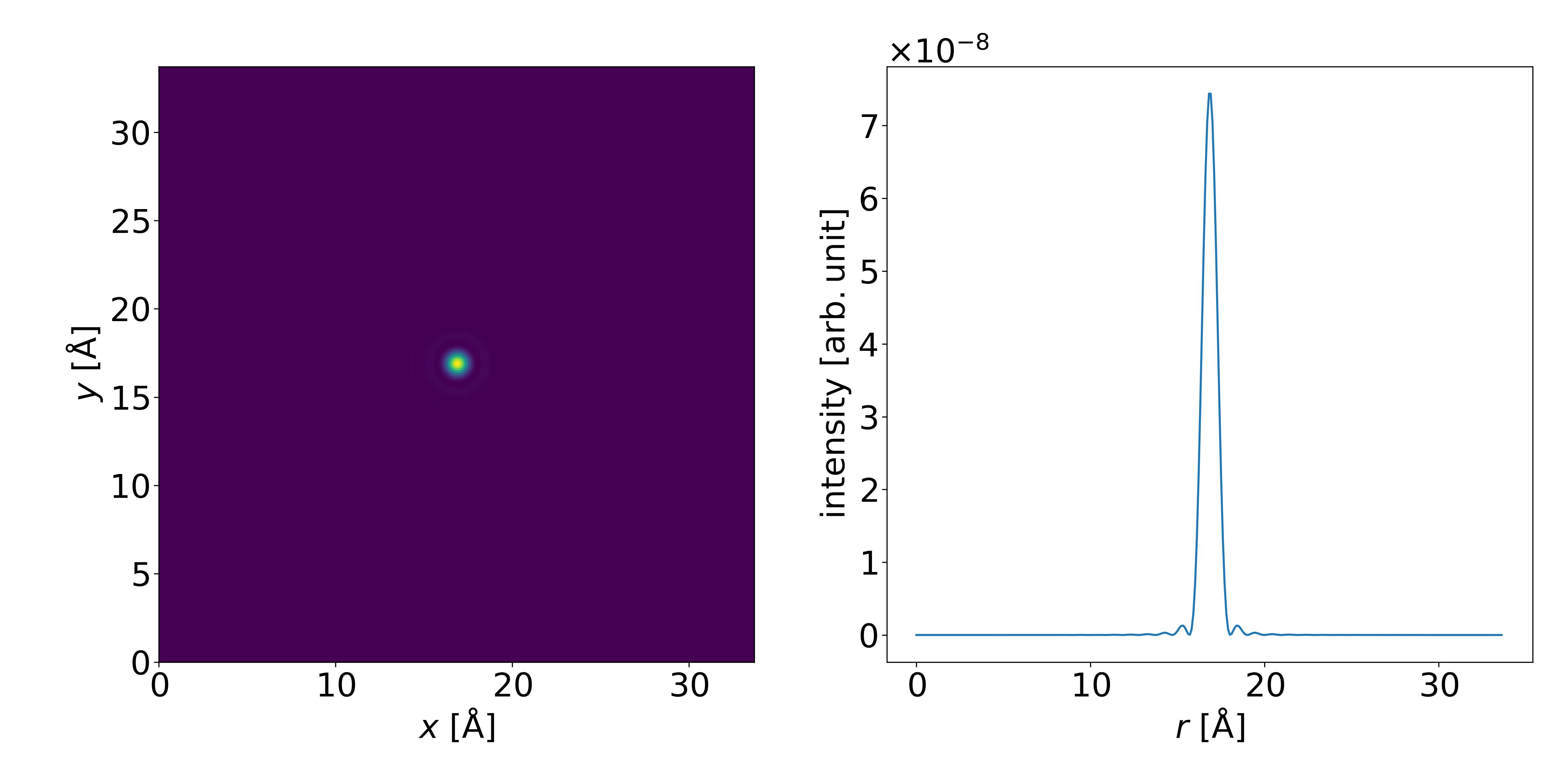}
            \caption{Simulated 2D intensity distribution of an aberration-free focused electron probe (left), along with its corresponding line profile (right). The probe was modeled using a convergence semi-angle of 13 mrad and an acceleration voltage of 200 kV.}
            \label{fig-2}
        \end{figure}
        
        \begin{figure}[h!]
            \centering
            \includegraphics[width=0.5\textwidth]{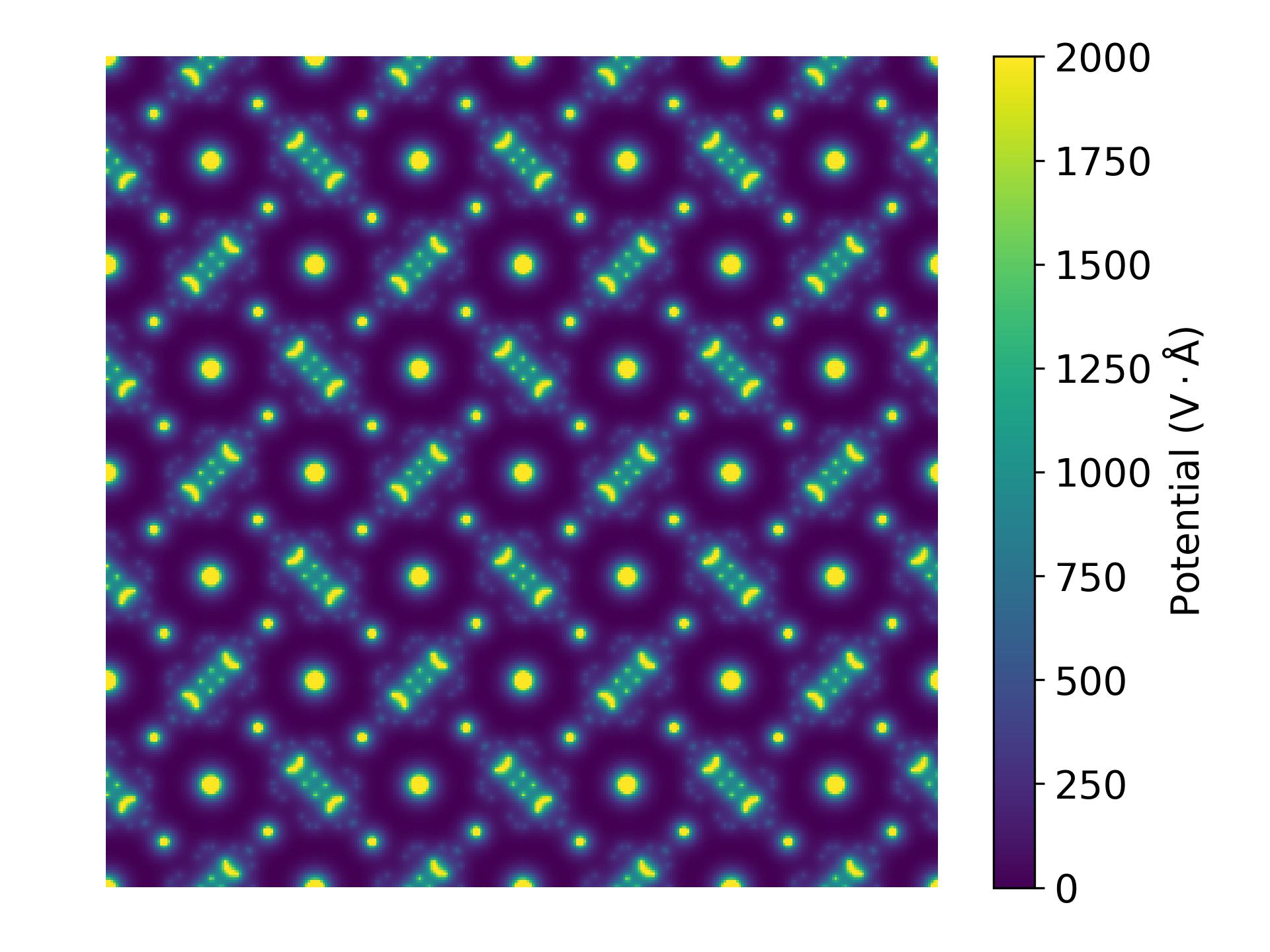}
            \caption{Vertical projection of the electrostatic potential of 2nm thick FAPbBr$_{3}$ used for the simulations. The colorbar reflects the values of the projected potential in V.$\text{\AA}$}
            \label{fig-3}
        \end{figure}
        
        \begin{figure*}[h!]
            \centering
            \includegraphics[width=1.0\textwidth]{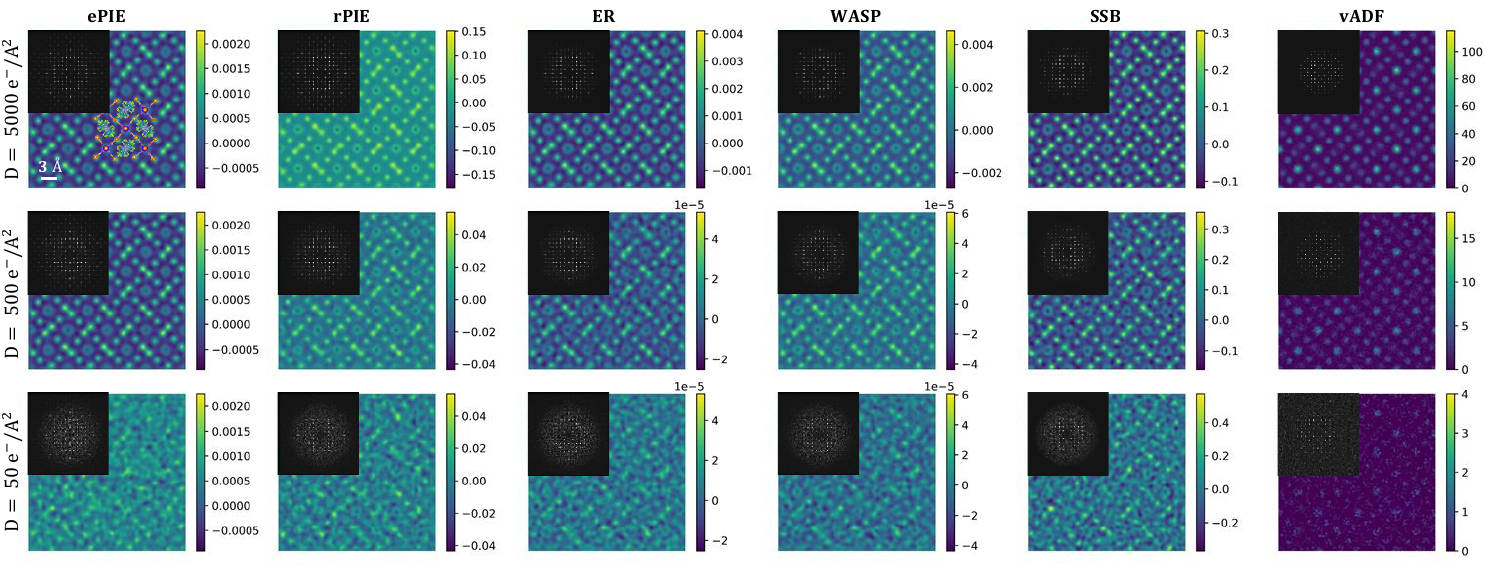}
            \caption{Ptychographic reconstructions and their corresponding Fourier transforms from a simulated 4D dataset of approximately 2 nm thick FAPbBr$_{3}$, alongside vADF images, are shown at varying electron doses. The colorbars represent phase values in radians for the ptychographic reconstructions and electron scattering intensities for the vADF images.}
            \label{fig-4}
        \end{figure*}
        
        A common criticism of SP, i.e. ePIE-like, algorithms is that they are time-consuming and become increasingly computationally expensive as the batch size or number of iterations increases. This is due to the sequential nature of the updates, where each step depends on the outcome of the previous one, preventing parallelization. 

        \begin{align}
            P_{j+1}(\vec{r}) &= \frac{\sum_{j} O_j^*(\vec{r}) \psi_j^{'}(\vec{r})}{\sum_{j} |O_j(\vec{r})|^2},
            \label{er-update-probe}\\
            O_{j+1}(\vec{r}) &= \frac{\sum_{j} P_j^*(\vec{r}) \psi_j'(\vec{r})}{\sum_{j} |P_j(\vec{r})|^2}.
            \label{er-update-object}
        \end{align}

        Batch processing algorithms such as ER \cite{thibault_high-resolution_2008, thibault_probe_2009} treat the whole data as part of each update. As such, they consist in calculating the exit waves for each probe position and then feeding them into the updated probe and object functions as described in equation \ref{er-update-probe} and \ref{er-update-object}. The ability to treat the entire batch of probe positions at once allows parallelizing the calculation of the exit waves and, as a consequence, lowering computation times compared to SP algorithms.
        
        Finally, the WASP algorithm \cite{maiden_wasp_2024} is a hybrid approach that combines the standard update function of rPIE with the batch processing strategy of ER. Integrating rPIE into the algorithm is believed to enhance the convergence speed of ER.

\section{Evaluation using simulated data}
\label{simulation results}
    
    \subsection{\textit{Performance of iterative methods and SSB at high and low electron doses}}
                
        \begin{figure*}[h]
            \centering
            \includegraphics[width=1\textwidth]{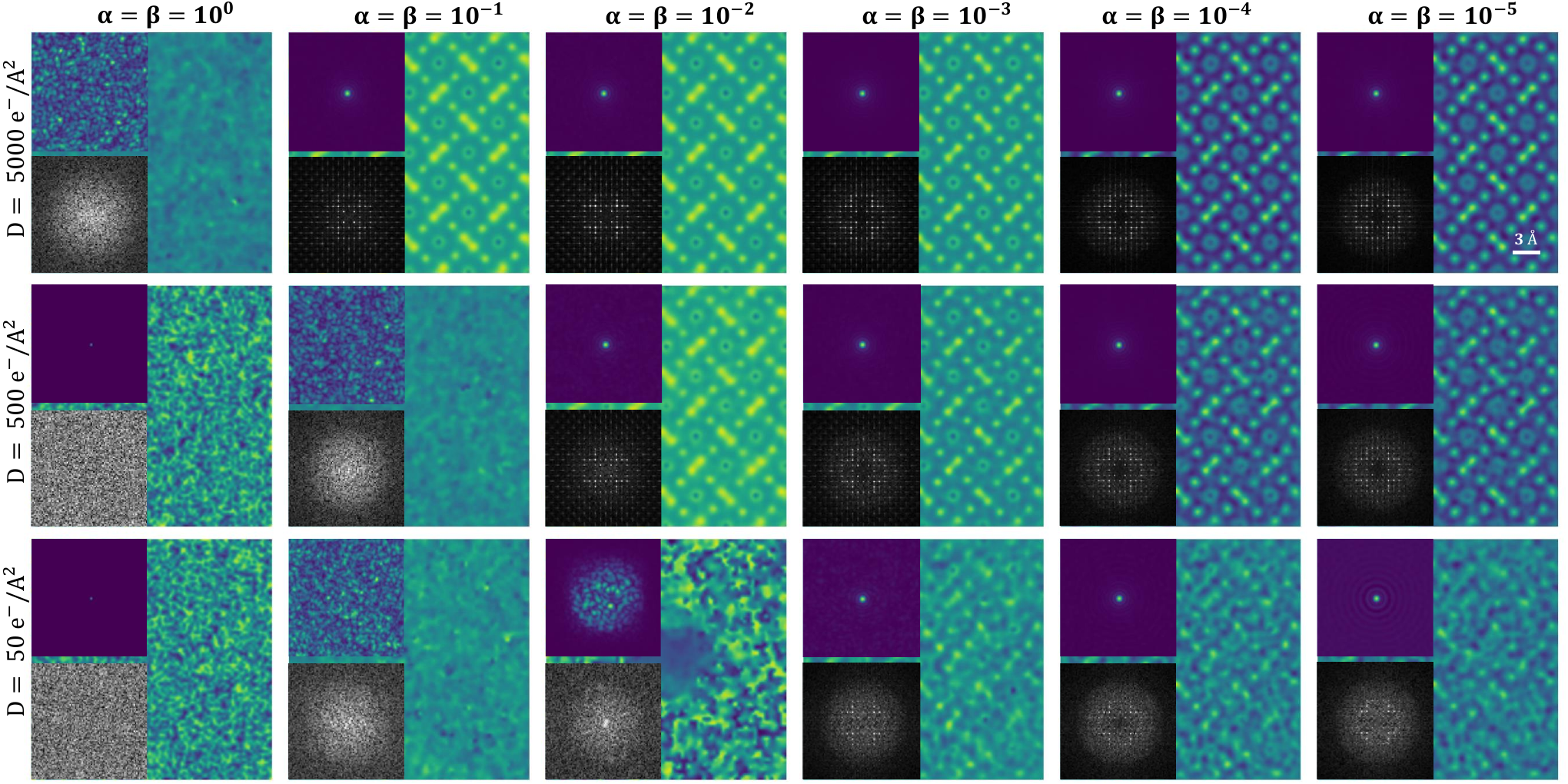}
            \caption{ePIE reconstructions at varying electron doses and update strengths $\alpha$ and $\beta$. For each condition, the reconstructed object's phase, the amplitude of its Fourier transform, and the amplitude of the reconstructed probe are shown side by side for direct comparison}
            \label{fig-5}
        \end{figure*}
        
        \begin{figure}[h]
            \centering
            \includegraphics[width=0.5\textwidth]{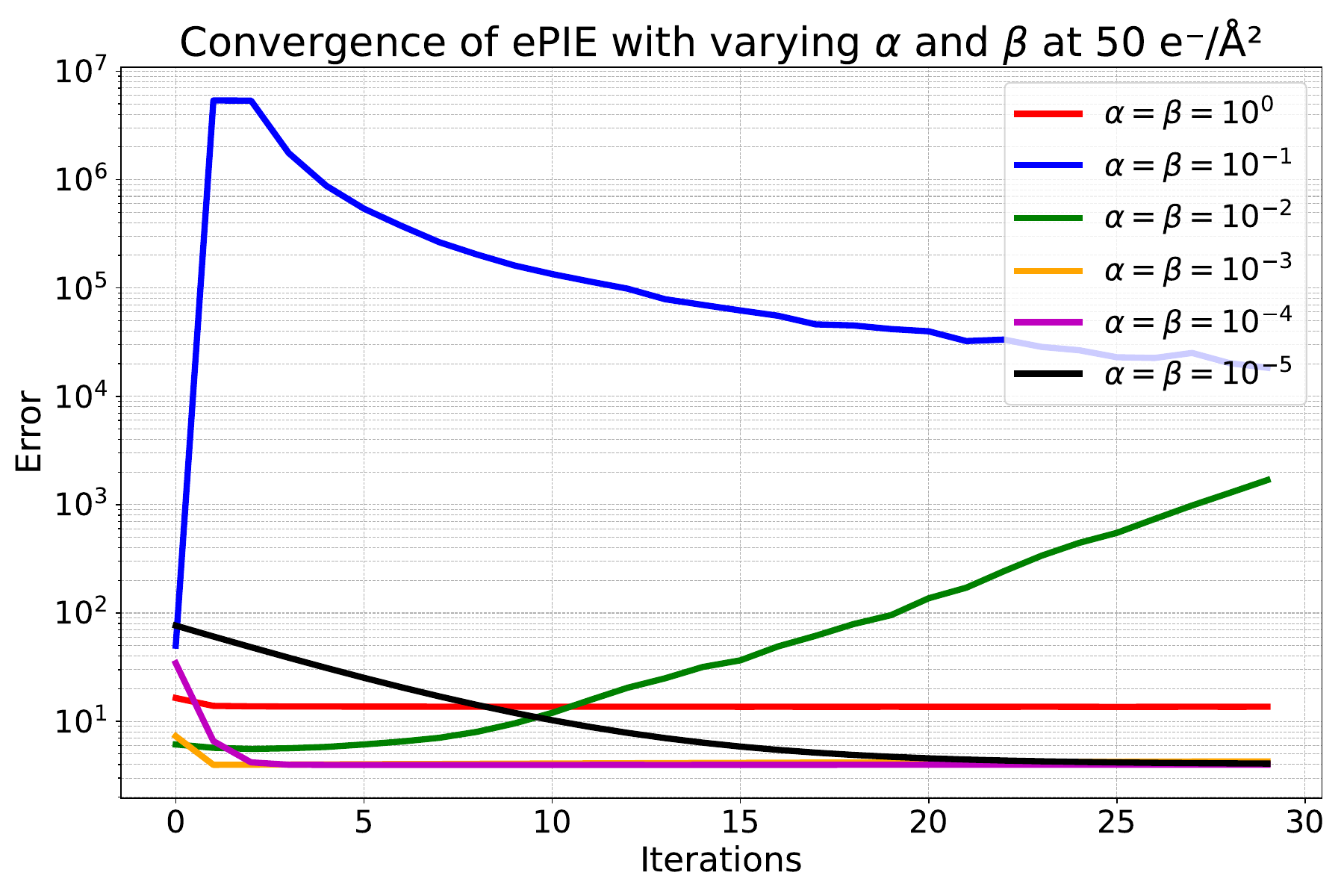}
            \caption{Error curves of ePIE at 50 $e^-/\text{\AA}^2$ for varying update strengths.}
            \label{fig-6}
        \end{figure}
        
        \begin{figure*}[h]
            \centering
            \includegraphics[width=1\textwidth]{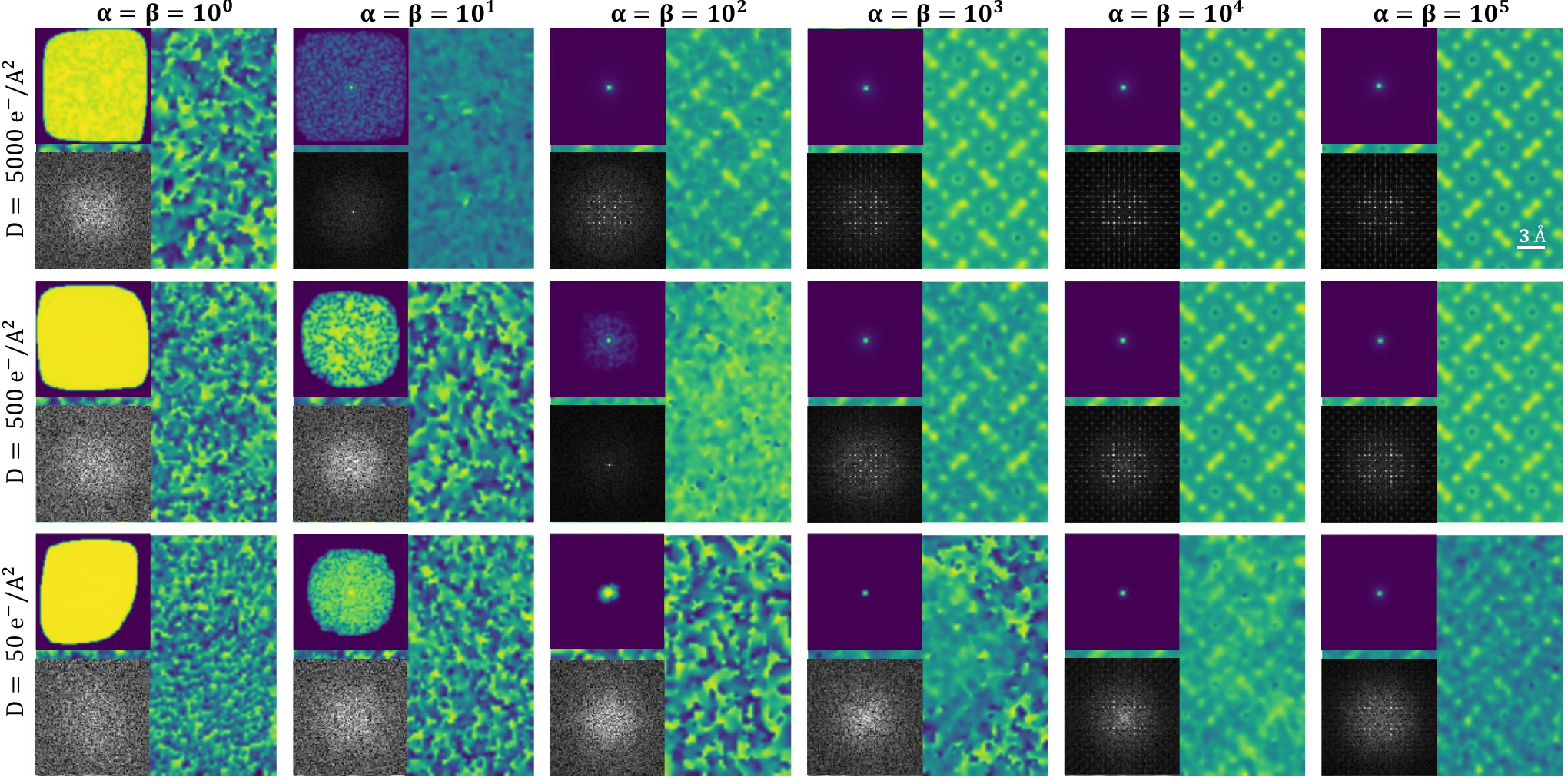} 
            \caption{rPIE reconstructions at varying electron doses and update strengths $\alpha$ and $\beta$. Each reconstruction is displayed alongside its corresponding FFT and reconstructed probe amplitude for direct comparison.}
            \label{fig-7}
        \end{figure*}
        
        \begin{figure}[h]
            \centering
            \includegraphics[width=0.48\textwidth]{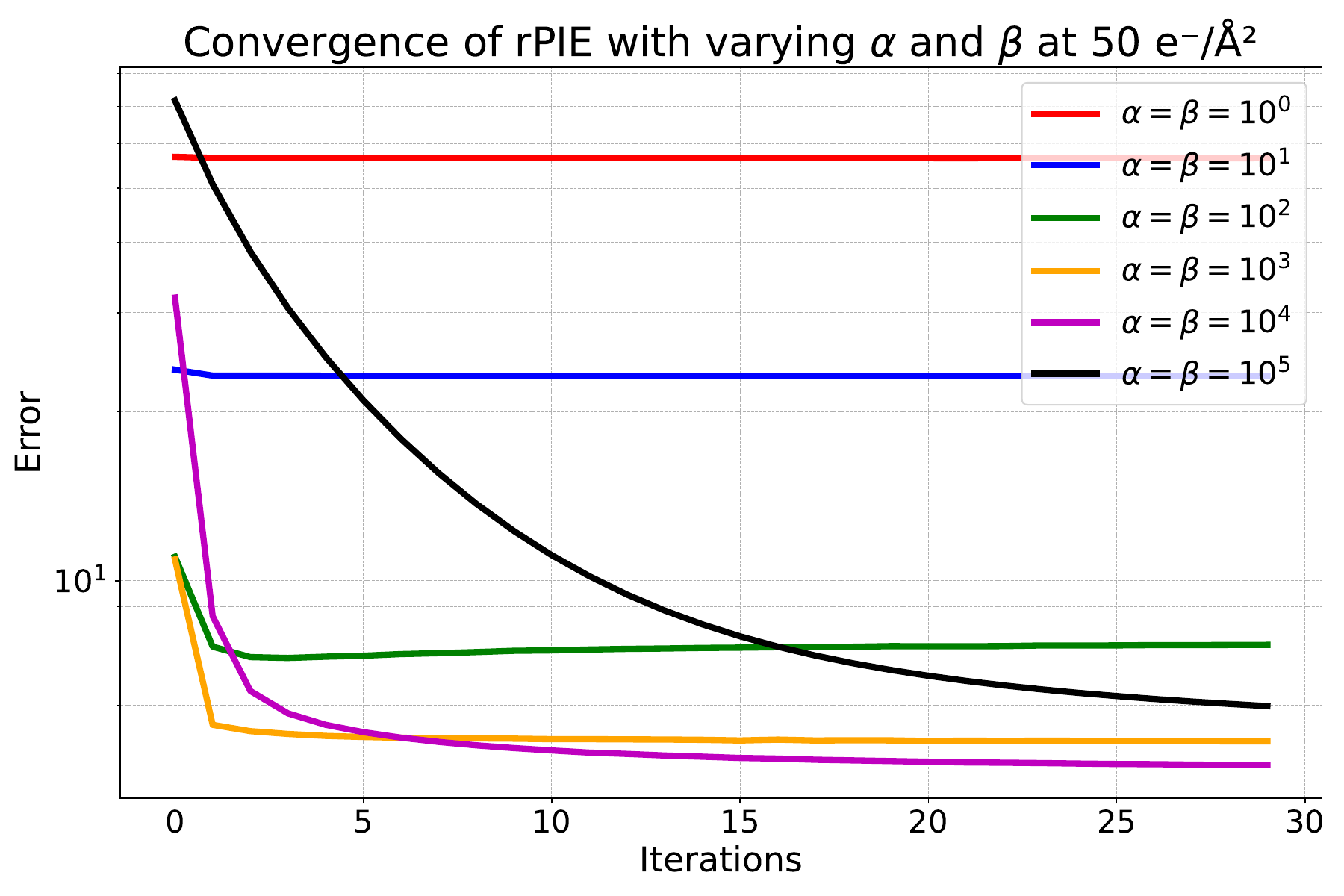}
            \caption{Error curves of rPIE at 50 $e^-/\text{\AA}^2$ for varying update strengths."}
            \label{fig-8}
        \end{figure}

        \begin{figure*}[h]
            \centering
            \includegraphics[width=1\textwidth]{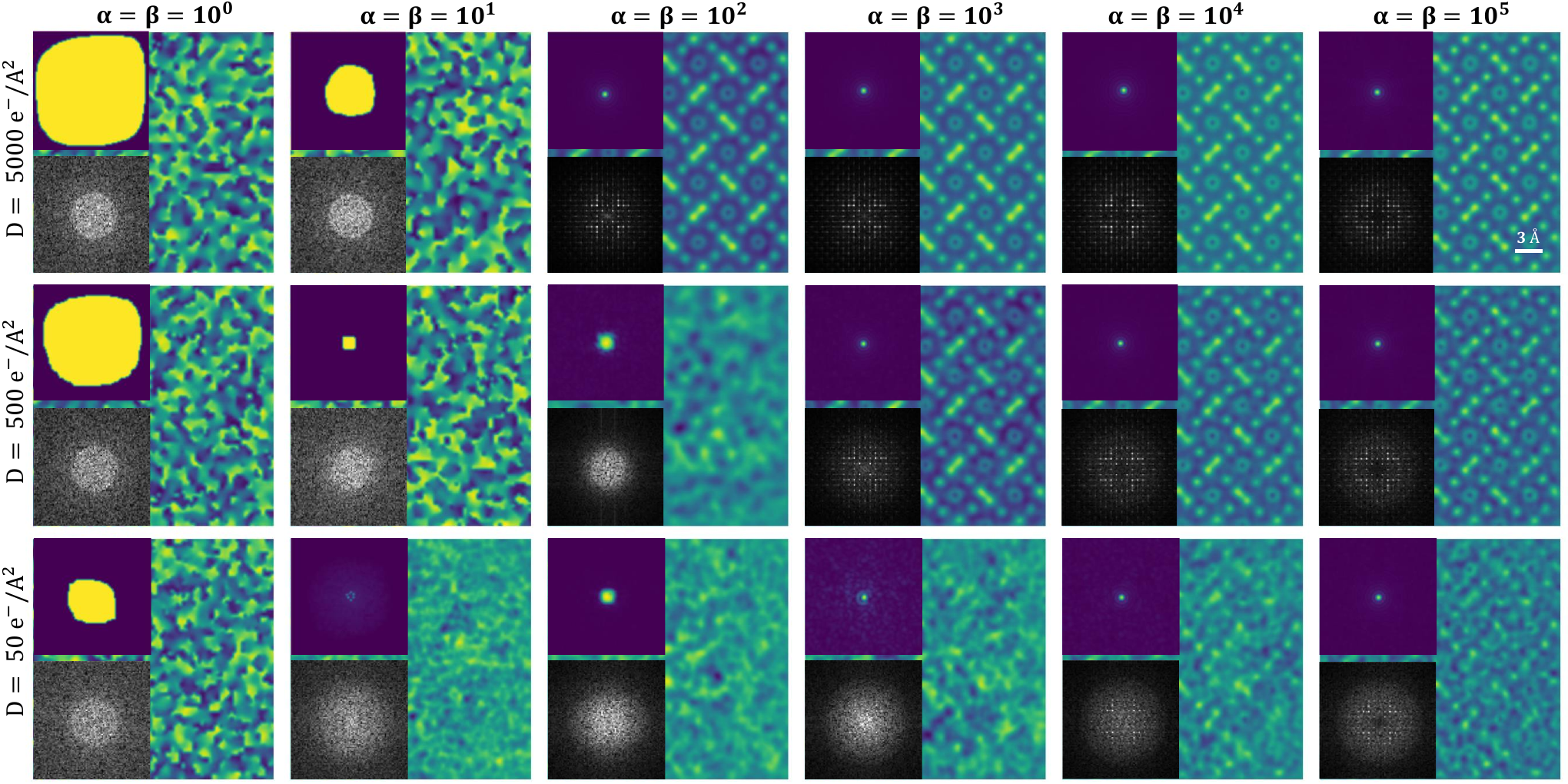}
            \caption{WASP reconstructions at varying electron doses and update strengths $\alpha$ and $\beta$. Each reconstruction is displayed alongside its corresponding FFT and reconstructed probe amplitude for direct comparison.}
            \label{fig-9}
        \end{figure*}
        
        \begin{figure}[h]
            \centering
            \includegraphics[width=0.48\textwidth]{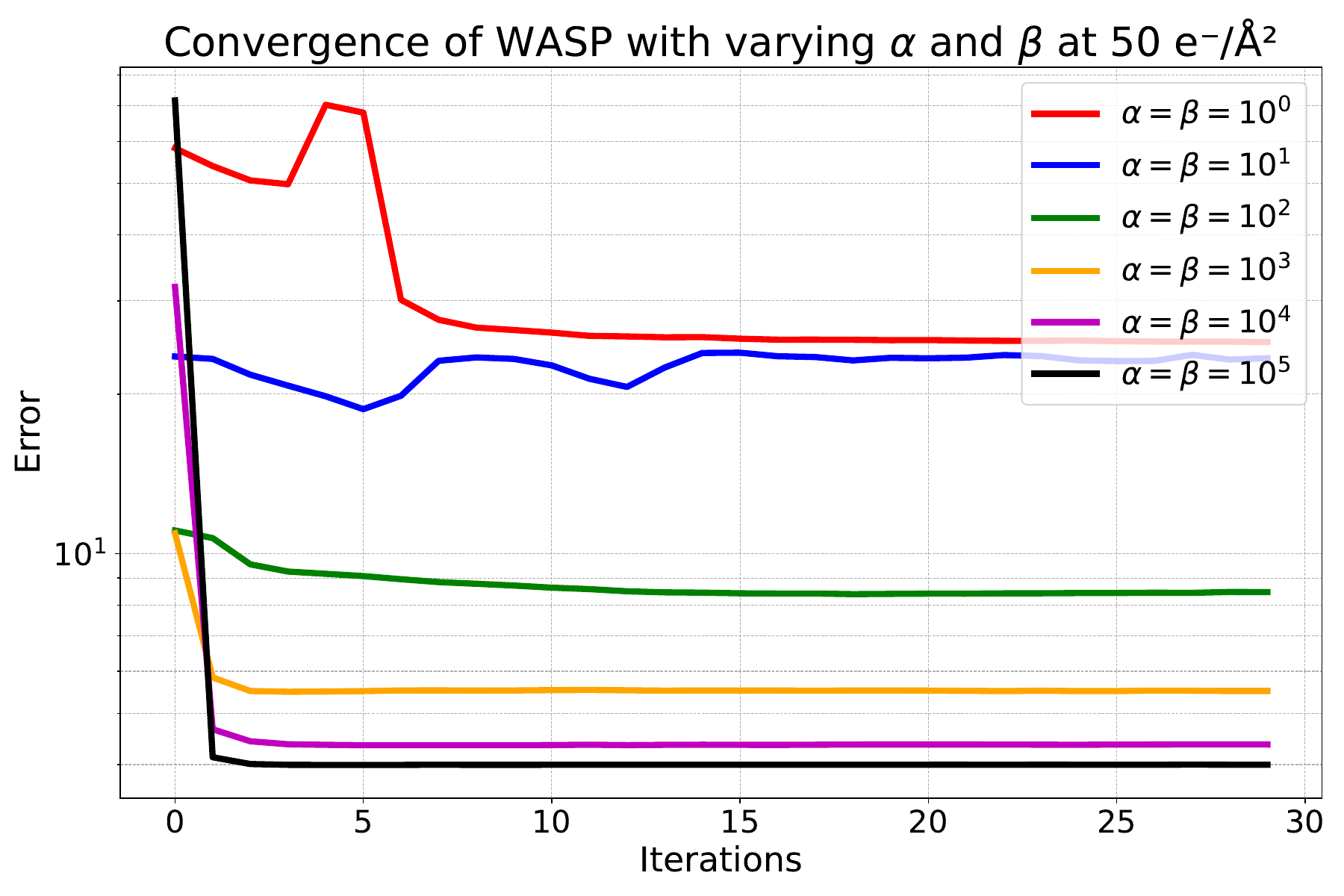}
            \caption{Error curves of WASP at 50 $e^-/\text{\AA}^2$ for varying update strengths.}
            \label{fig-10}
        \end{figure}

        To evaluate the performance of the ptychographic algorithms described above, we simulated a 4D dataset using the abTEM software \cite{madsen_abtem_2021}. The test sample consisted of a 2 nm thick crystal of formamidinium lead bromide (FAPbBr${_3}$), a hybrid organic–inorganic perovskite, where FA (formamidinium) is the organic cation with the chemical formula CH${5}$N$_{2}^{+}$. Such hybrid organic-inorganic perovskites are considered promising semiconductors for several optoelectronic technologies, thanks to their tunable bandgap and high carrier mobility among other properties \cite{huang_solar-driven_2020, dey_state_2021}.
        
        The simulation used an acceleration voltage of 200 kV, a convergence angle of 13 mrad, and a focused aberration free probe. Sufficient information redundancy is a key factor for a successful iterative ptychography reconstruction, and is achieved via the overlap of the areas illuminated at adjacent probe positions. Typically, an overlap of approximately 60-80\% is sufficient \cite{bunk_influence_2008}. In this case, a step size of approximately 20 pm was employed to ensure an overlap of about 80\%. The propagation of the electron wave through the specimen was performed under the multislice approximation. \cite{cowley_scattering_1957, goodman_numerical_1974, ishizuka_new_1977}. Owing to the dominant use of low-angle scattering, and hence the minor role of thermal diffuse scattering in this application of electron ptychography, lattice vibrations were not considered. Figure \ref{fig-3} is the resulting projected potential of the 2 nm thick FAPbBr$_{3}$ specimen. The probe intensity is given in figure \ref{fig-2}. Dose-limited, and hence noisy, data is obtained from the simulated 4D dataset by applying Poisson statistics.
        
        Figure \ref{fig-4} shows simulated ptychographic reconstructions obtained with the ePIE, rPIE, ER, WASP, and SSB algorithms, together with a virtual annular dark-field (vADF) image generated by integrating the intensity over the angular range of 14–32 mrad. The pixelated detector used here has a maximum collection angle of 32 mrad. The ePIE, rPIE, ER, and WASP implementations were adapted from Ref. \cite{andrew_maiden_ptychography_2025} and translated into Python, while SSB reconstructions were carried out using pyPtychoSTEM \cite{noauthor_pyptychostem_2024}. For the update coefficients, we used $\alpha = \beta = 10^{-4}$ in ePIE and $\alpha = \beta = 10^{5}$ in rPIE and WASP. These values were selected empirically from parameter sets that yielded convergence. In the following section, we discuss the influence of different coefficient choices on algorithmic convergence and the resulting phase-object reconstructions.

        A total of 30 iterations was employed for the iterative algorithms to ensure sufficient convergence of the reconstruction while balancing computational efficiency. This number was chosen based on prior experience and convergence behavior observed in similar datasets, where further iterations showed diminishing improvements.
        
        The atomic sites of FAPbBr$_{3}$ are evident at the highest dose 5000 $e^-/\text{\AA}^2$ in both the ptychographic reconstructions and the vADF; however, the positions of the organic components in the structure are more discernible in the ptychographic reconstructions, whereas only the heavier Pb and Br columns are visible in the ADF. At 500 $e^-/\text{\AA}^2$, both the vADF and the ptychographic reconstructions exhibit increased noise, although the contrast from the ptychographic reconstructions is better compared to the vADF. At the lowest electron dose used here, the noise level becomes more significant in both imaging modes; the lower frequencies still appear more readily observable in the ptychographic reconstructions than in the vADF image.
        
        Notably, this analysis assumes an ideal case in which the sample is not on any support. This differs from experimental conditions, where the sample is often deposited on amorphous carbon or grown on a substrate. Furthermore, contrast reversals observed on the Pb columns in the ptychographic reconstructions are primarily caused by the strong projected potentials at these sites. These contrast reversals can be corrected post-reconstruction using the phase offset method \cite{hofer_phase_2024}.
    
    \subsection{\textit{The effect of the update coefficients on iterative methods}}
        
        Finding the optimal update coefficients for ptychographic reconstructions can be challenging, as excessive modifications of the probe and object functions often lead to a divergence of the reconstruction process, while smaller ones may cause the algorithms to become trapped in local minima. In this section, we investigate the effect of the update strengths at high and low electron doses, for both the extended and the regularized version of PIE, as well as WASP.
        
        Figure \ref{fig-5} depicts ePIE reconstructions based on the simulated 4D dataset presented above, using different update strengths and doses. At the highest dose of 5000 $e^-/\text{\AA}^2$ ePIE fails to converge to an image showing the lattice within 30 iterations using $\alpha$ = $\beta$ = 1. In addition, the reconstructed probe is clearly incorrect, as evidenced by the probe amplitude displayed alongside the reconstruction in this case. Relatively good object reconstructions are on the other hand achieved for $\alpha = \beta = 10^{-1}$, $10^{-2}$, and $10^{-3}$, however, the square root of the Fourier transform amplitudes shown as insets reveal artefactual low-frequency components, while the probe's amplitude is being correctly reconstructed at the same values. At the two lowest update values $\alpha$ = $\beta$ = 10$^{-4}$ and $\alpha$ = $\beta$ = 10$^{-5}$, artifact free reconstructions are obtained as depicted from their respective FFT as well as correctly recovered probes amplitudes. At the intermediate dose of 500 $e^-/\text{\AA}^2$ and for the two highest update strength values, the reconstruction fails to capture the specimen's structure, while the same artefactual low-frequency components are still present at $\alpha$ = $\beta$ = 10$^{-2}$ and 10$^{-3}$. Similarly to the previous case, the reconstruction of the  probe's amplitude is clearly wrong at the two highest coefficients as depicted by the figure, while both the object and probe's amplitude are correctly recovered at the two lowest values. At the lowest electron dose of 50 $e^-/\text{\AA}^2$, the algorithm fails to converge at the three highest values of $\alpha$ and $\beta$. Finally, for all three doses, improved reconstructions of both the probe and object functions are achieved at values of 10$^{-3}$, 10$^{-4}$, and 10$^{-5}$.
        
        The error curves for the different update strengths, given a dose of 50 $e^-/\text{\AA}^2$, are given in figure \ref{fig-6}. Note that the object reconstruction begins from a uniform initial guess, with the phase set to zero and the amplitude set to one. As for the probe it is assumed to be focused and aberration-free. The sum squared error (SSE) is calculated according to equation \ref{SSE} and provided as a function of the number of iterations. First, we observe that, while the error curve shows the error decreasing after the first few iterations at the high values $\alpha$ = $\beta$ = 1, the corresponding result does not display useful structural information. This is likely an indication that the algorithm is reaching a local minimum and thus providing an incorrect solution for the object and/or probe functions. At $\alpha$ = $\beta$ = 10$^{-1}$, an increase of the error occurs at the first few iterations, leading to a slow convergence process conserving a high error value for the rest of the calculation, and thus possibly reaching another, different, local minimum. At $\alpha$ = $\beta$ = 10$^{-2}$, the algorithm exhibits a clear diverging behavior. Finally, the error curves for $\alpha$ = $\beta$ = 10$^{-3}$, 10$^{-4}$ and 10$^{-5}$ show the desired convergent behavior, and the resulting images indeed show the lattice. It is worth noting that, for $\alpha$ = $\beta$ = 10$^{-4}$ and 10$^{-5}$, the initial error starts increasing in the early stage of the reconstruction, but ends up reaching the same minimum error as the threshold value obtained with $\alpha$ = $\beta$ = 10$^{-3}$.
        
        The same investigation is next performed with rPIE. Importantly, here, the update coefficients are in the denominator, hence higher values imply a slower modification of the probe and object functions, opposite to the behavior with ePIE. From figure \ref{fig-7} we observe that, at the highest dose of 5000 $e^-/\text{\AA}^2$, the reconstruction does not lead to useful structural information for $\alpha$ = $\beta$ = 1, whereas successful reconstructions are achievable for higher coefficients. Similar behavior is also observed for the probe's amplitude, which is clearly wrong at the two lowest update coefficients values as shown in the figure.
        
        At 500 $e^-/\text{\AA}^2$, the process fails at the three lowest values of $\alpha$ and $\beta$ while correct convergence is achieved for higher values. At the lowest electron dose of 50 $e^-/\text{\AA}^2$, for the four lowest update coefficients neither the lower or higher frequencies of the object are reconstructed, even though the error curves provided in figure \ref{fig-8}show a convergent behavior. This, again, indicates an incorrect solution being reached by the algorithm. Finally, at $\alpha$ = $\beta$ = 10$^{4}$, a first correct transfer of the specimen's frequencies occurs, as evidenced by the FFT, and the overall contrast of the images further improves for $\alpha$ = $\beta$ = 10$^{5}$. For the amplitude of the probe, an inaccurate reconstruction is observed at the three lowest coefficients values, whereas a more correct probe distribution is achieved at the highest values. Similarly to the ePIE case, the error curves show the expected convergence, although the initial SSE values increase as the coefficients get larger. The lowest error is achieved for a threshold set of values $\alpha$ = $\beta$ = 10$^{4}$.
        
        Figure \ref{fig-9} displays results from reconstructions using the WASP algorithm. In general, they show similar features and dose/update strength dependence as ePIE and rPIE. First, at the highest dose, the algorithm fails to converge with the smallest coefficients but converges without artifacts with all others employed. At 500 $e^-/\text{\AA}^2$, and at the three lowest coefficient values, neither the lower nor the higher frequencies are reconstructed, while the process still converges at the highest values. Finally, at 50 $e^-/\text{\AA}^2$, the calculation starts converging at $\alpha$ = $\beta$ = 10$^{3}$. Figure \ref{fig-10} depicts the error curves for the different coefficients, at the lowest introduced electron dose, like in the previous cases. As a first observation, all the curves show convergence for the process, though the lowest converged values is only achievable for the highest two update strengths. Additionally, all the successful reconstructions are free from artifacts. This is interesting, as this indicates that WASP is potentially less susceptible to incorrect solutions and local minima than ePIE and rPIE.
        
        Overall, this simulation study suggests that under low-dose conditions, iterative ptychographic algorithms require small update coefficients for successful reconstructions preferably $\alpha = \beta = 10^{-3}$ for ePIE, and $\alpha = \beta = 10^{4}$ for rPIE and WASP. Those values are however specific to the investigated material case and should be adapted to the particular experimental configuration and specimen. Nevertheless, the general trends found for the dependencies on dose and update coefficients, for all investigated algorithms, can reasonably be expected to remain for most applications.

\section{\textit{Evaluation using experimental data}}
    
    \begin{figure}[h!]
        \centering
        \includegraphics[width=0.48\textwidth]{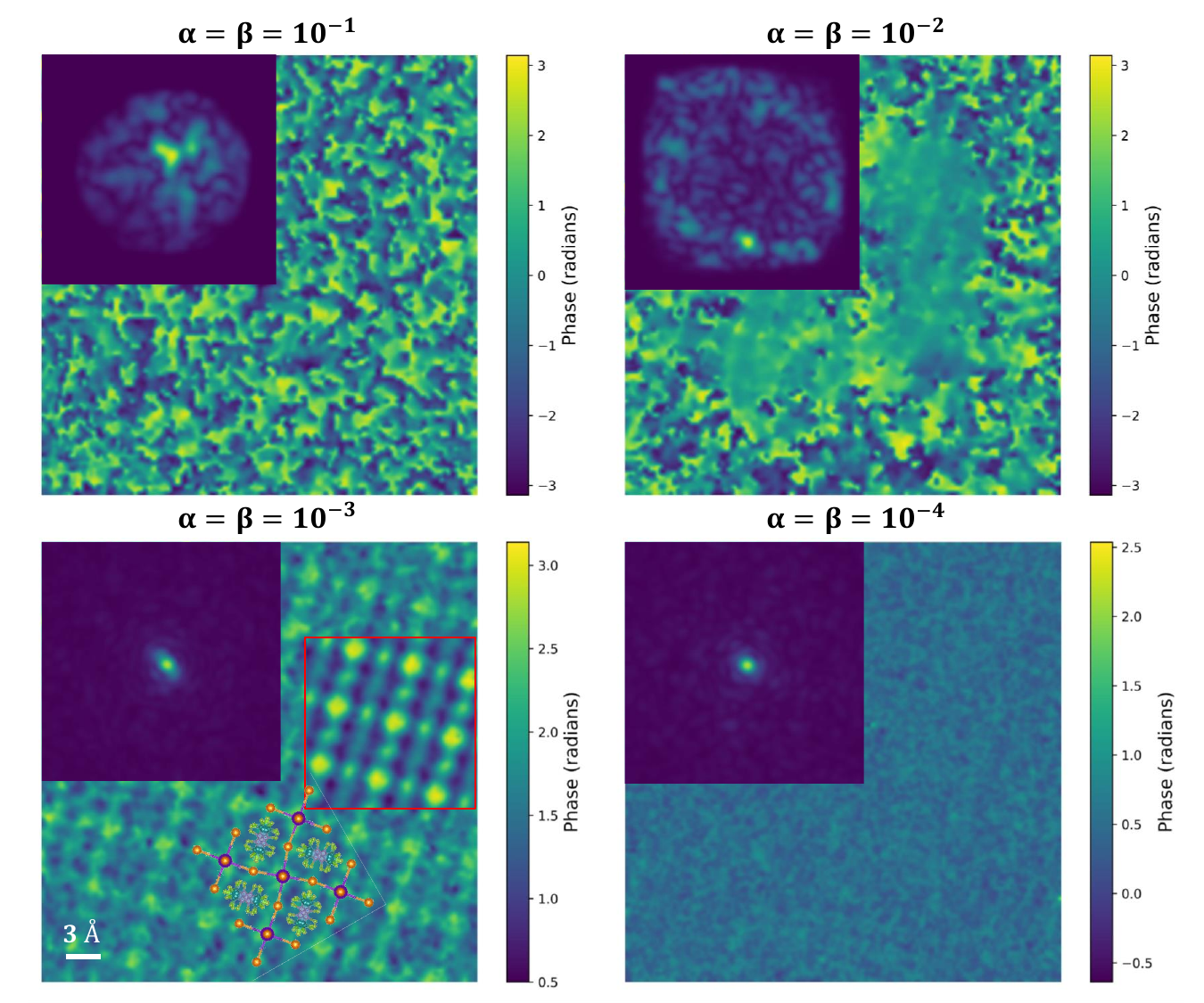}
        \caption{ePIE reconstructions of FaPbBr$_{3}$ at varying update strengths using an experimental dataset acquired with a dose of 50 $e^-/\text{\AA}$, shown with the corresponding reconstructed probe amplitudes. The 4D-STEM data of the nanocrystal (NC) depicted in the reconstruction for $\alpha = \beta = 10^{-3}$ was averaged using template matching and is presented alongside the reconstruction.}
        \label{fig-11}
    \end{figure}
    
    \begin{figure}[h]
        \centering
        \includegraphics[width=0.48\textwidth]{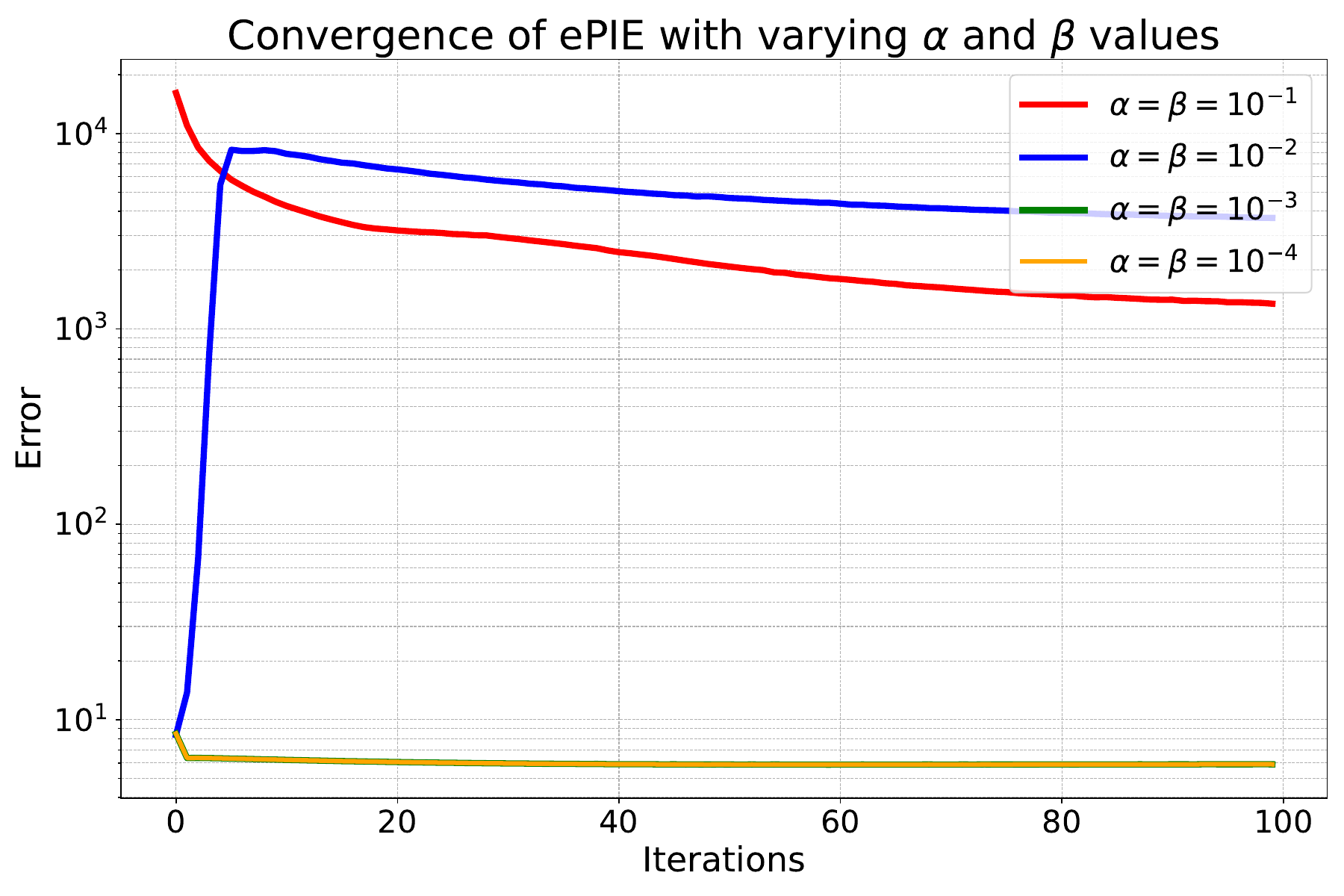}
        \caption{Error curves for ePIE reconstructions at varying update strengths using an experimental dataset acquired with a dose of 50 e$^-$/Å$^2$.}
        \label{fig-12}
    \end{figure}
    
    \begin{figure}[h!]
        \centering
        \includegraphics[width=0.48\textwidth]{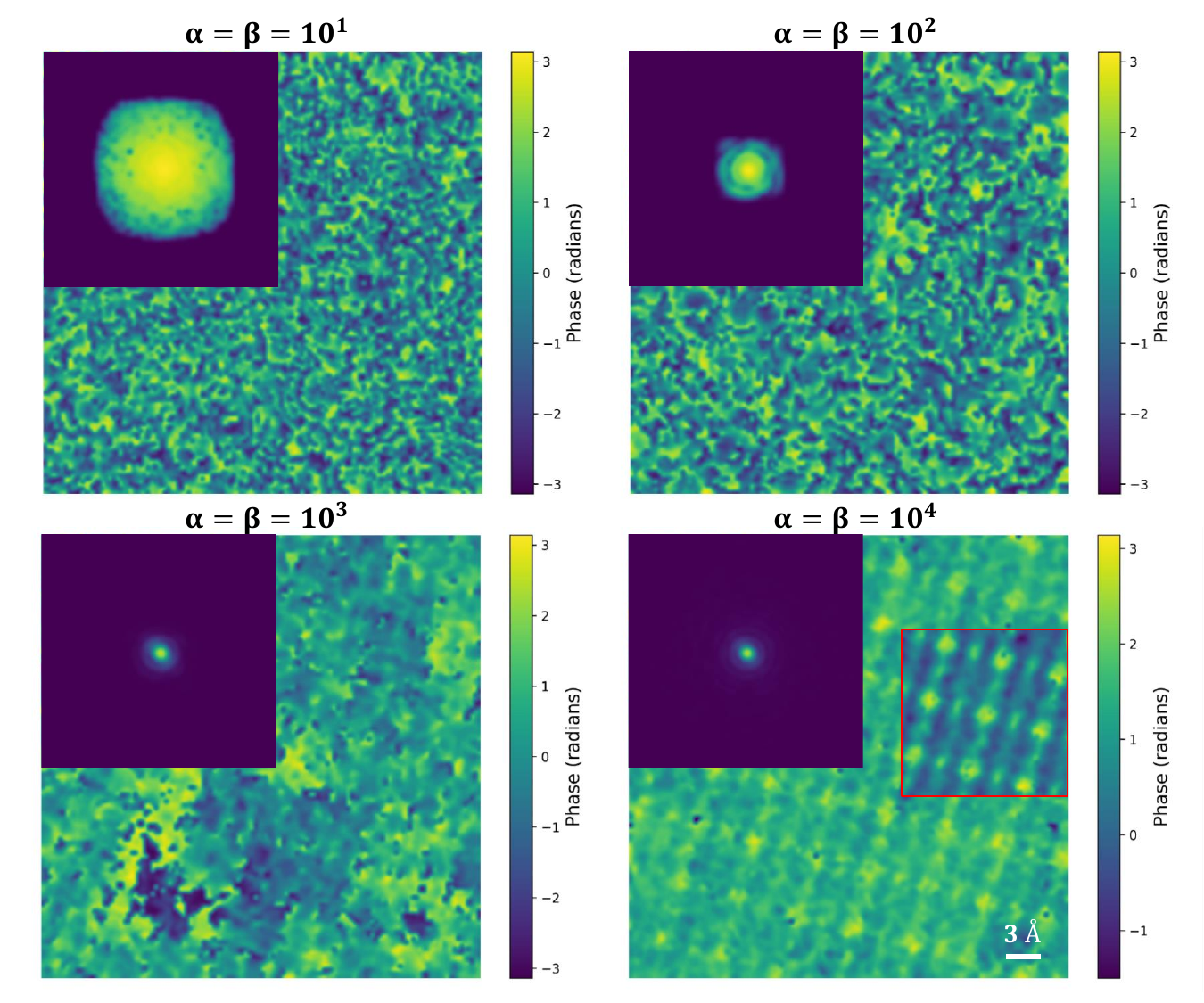}
        \caption{rPIE reconstructions of FaPbBr$_{3}$ at varying update strengths using an experimental dataset acquired with a dose of 50 $e^-/\text{\AA}$, shown with the corresponding reconstructed probe amplitudes. The 4D-STEM data of the nanocrystal (NC) depicted in the reconstruction for $\alpha = \beta = 10^{4}$ was averaged using template matching and is presented alongside the reconstruction.}
        \label{fig-13}
    \end{figure}
    
    \begin{figure}[h!]
        \centering
        \includegraphics[width=0.48\textwidth]{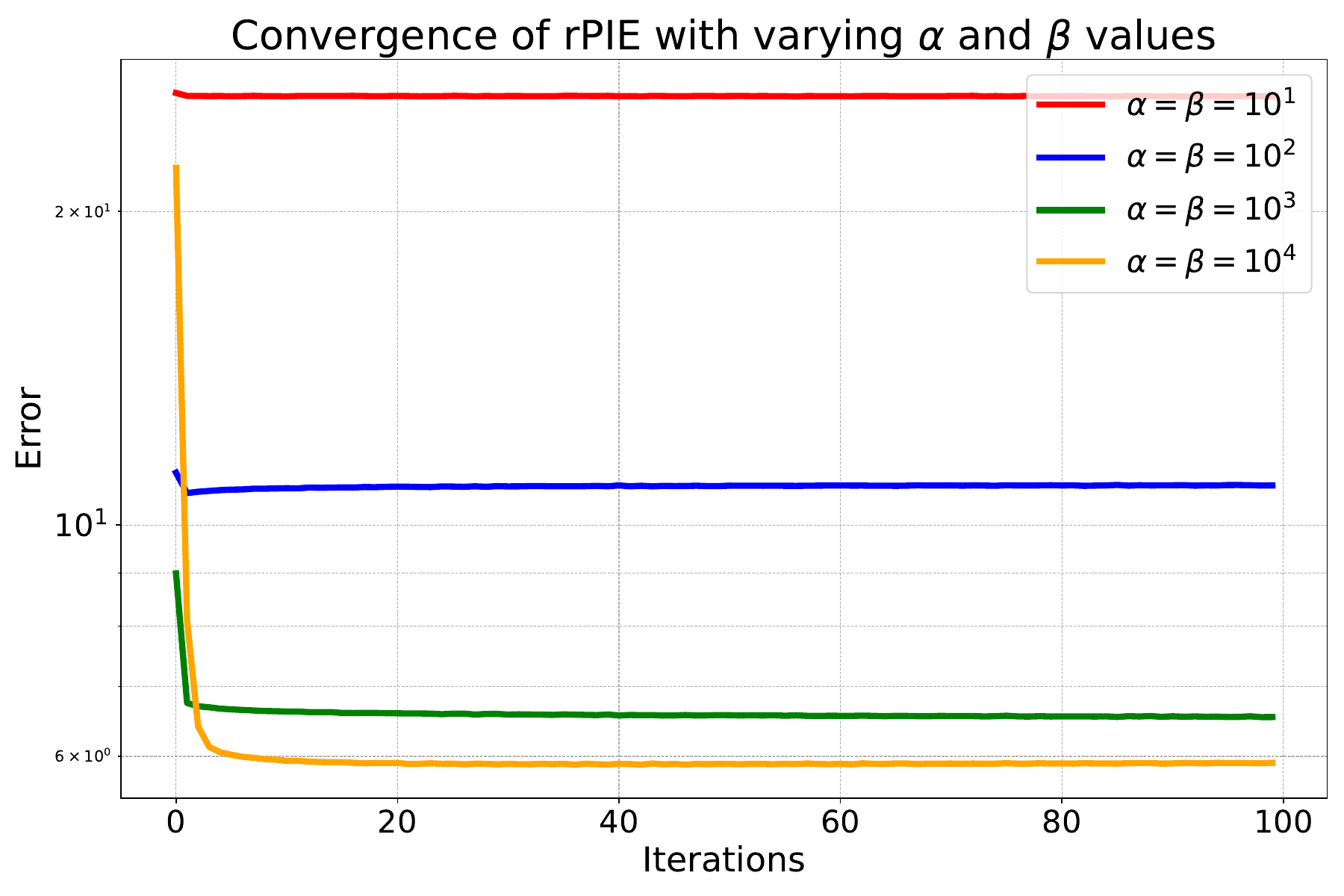}
        \caption{Error curves for rPIE reconstructions at varying update strengths using an experimental dataset acquired with a dose of 50 e$^-$/Å$^2$.}
        \label{fig-14}
    \end{figure}
    
    \begin{figure}[h!]
        \centering
        \includegraphics[width=0.48\textwidth]{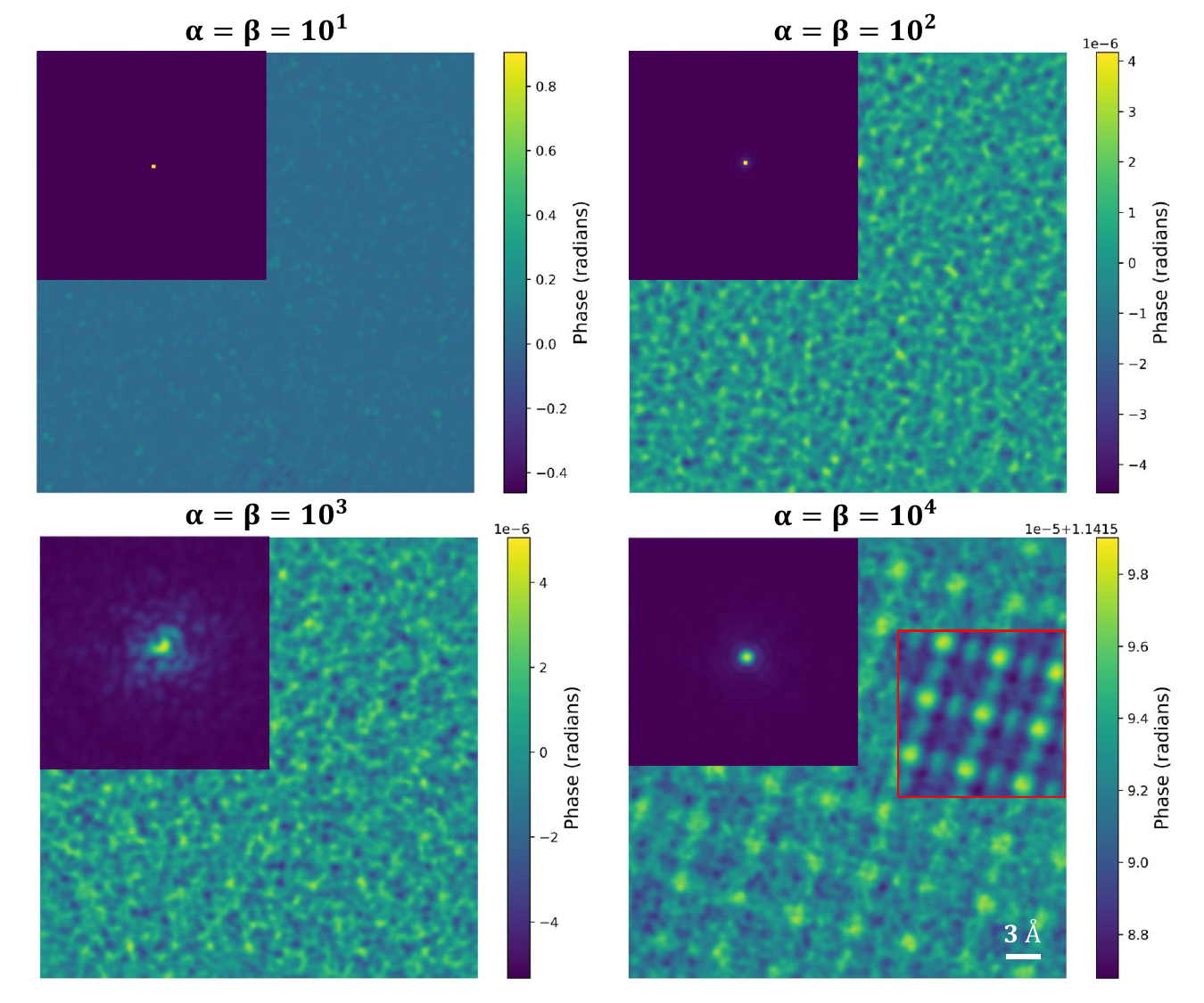}
        \caption{WASP reconstructions of FaPbBr$_{3}$ at varying update strengths using an experimental dataset acquired with a dose of 50 $e^-/\text{\AA}$, shown with the corresponding reconstructed probe amplitudes. The 4D-STEM data of the nanocrystal (NC) depicted in the reconstruction for $\alpha = \beta = 10^{4}$ was averaged using template matching and is presented alongside the reconstruction.}
        \label{fig-15}
    \end{figure}
    
    \begin{figure}[h!]
        \centering
        \includegraphics[width=0.48\textwidth]{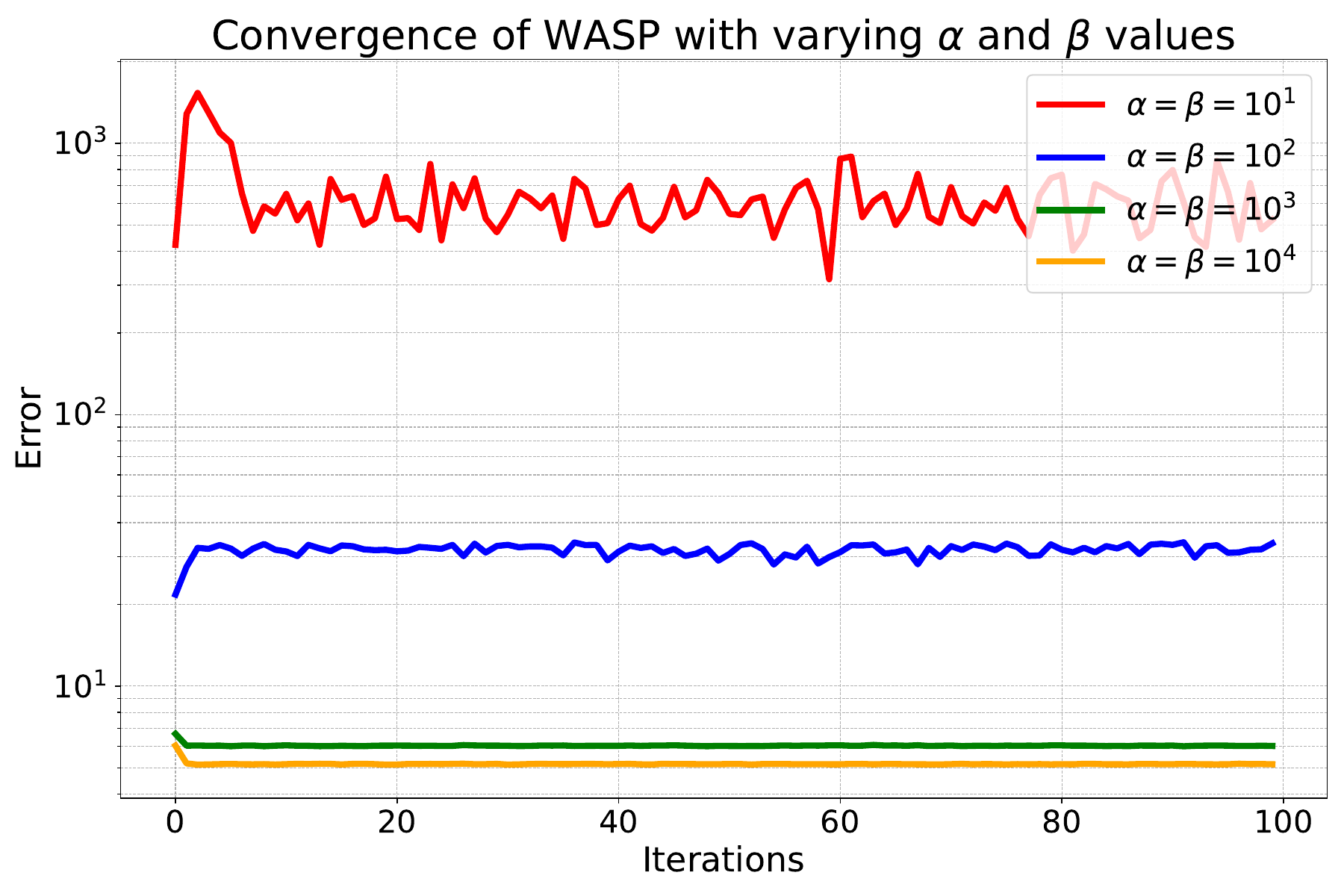}
        \caption{Error curves for WASP reconstructions at varying update strengths using an experimental dataset acquired with a dose of 50 e$^-$/Å$^2$.}
        \label{fig-16}
    \end{figure}
    
    To validate the simulation study presented in section \ref{simulation results}, a 4D dataset was experimentally obtained from a FAPbBr$_{3}$ nanocrystal, with a thickness approximately equal to 10 nm. The measurements were conducted with an acceleration voltage of 200 kV and a semi-convergence angle of 13 mrad, similar to the parameters chosen in the simulation study. The dataset was acquired using a Timepix3 camera \cite{jannis_event_2022, poikela_timepix3_2014}, with a dwell time of 1 microsecond. The electron dose was approximately 50 $e^-/\text{\AA}^2$. The scan window encompassed 2048$\times$2048 probe positions spread across an area of 10$\times$10 nm$^{2}$, and the CBED patterns were binned by a factor of four to 64$\times$64.
    
    Figure \ref{fig-11} presents ePIE reconstructions performed on a portion of the scanned surface using 100 iterations and varying update strengths. The optimal reconstruction is achieved with $\alpha = \beta = 10^{-3}$, as corroborated by the error curves in Figure \ref{fig-12}. Although higher update values of $10^{-1}$ and $10^{-2}$ exhibit convergent trends, they maintain significantly larger errors throughout the process. Notably, an initial increase in error occurs within the first few iterations for $\alpha = \beta = 10^{-2}$, and the resulting reconstruction suggests convergence to a local minimum, indicating an incorrect solution. Conversely, at the lower update setting of $\alpha = \beta = 10^{-4}$, the error curve converges to a minimum similar to the $10^{-3}$ case; however, after 100 iterations, the reconstruction still fails to accurately estimate the object function. This suggests that the algorithm is converging more slowly toward the correct object structure and likely requires additional iterations to fully recover it. Additionally, the initial error for this case is slightly higher than that of the $10^{-3}$ setting.

    Figure \ref{fig-13} illustrates rPIE reconstructions using the same dataset. With update coefficients set to $\alpha = \beta = 10^{4}$, both the probe and object functions are successfully retrieved, and the algorithm exhibits stable convergence toward a minimal error. As shown in Figure \ref{fig-14}, lower update coefficients of $\alpha = \beta = 10^{1}$, $10^{2}$, and $10^{3}$ still produce convergence, but the algorithm settles at notably higher error values after 100 iterations. This suggests entrapment in erroneous local minima, clearly visible in the reconstructions of Figure \ref{fig-13}.

    Figure \ref{fig-15} displays WASP reconstructions performed on the same dataset with various update coefficients. The most accurate probe and object reconstructions correspond to $\alpha = \beta = 10^{4}$, consistent with the minimum error shown in Figure \ref{fig-16}. Lower update values of $\alpha = \beta = 10^{1}$, $10^{2}$, and $10^{3}$ yield higher minimum errors and less precise reconstructions, as also demonstrated in Figure \ref{fig-15}.

\section{Conclusion}
    
    Since the introduction of the original PIE algorithm \cite{rodenburg_phase_2004}, several other iterative methods have emerged, including ePIE \cite{maiden_improved_2009}, rPIE \cite{maiden_further_2017}, WASP \cite{maiden_wasp_2024}, and ER \cite{thibault_high-resolution_2008}, among others. The ePIE and rPIE solutions, which are sequential projection algorithms, are similar in their approach to updating the probe and object functions and differ in the used regularizer. WASP, conversely, relies on hybrid processing: it adopts the structural framework of the conventional ER method while incorporating an ePIE-like update function as input.
    
    SP and HP algorithm types require appropriate update coefficients to ensure convergence, particularly at low electron doses. Through the analysis of both simulated and experimental 4D datasets of a thin FAPbBr$_{3}$ specimen, we conclude that ePIE, rPIE, and WASP necessitate relatively small steps to facilitate convergence. Furthermore, we note that the initial reconstruction  error increases when the update coefficients are reduced below a certain threshold. These observations underscore the importance of carefully selecting appropriate parameters to achieve convergence and accurate reconstructions, particularly under low electron dose conditions. They also emphasize the existence of critical threshold values, beyond which the algorithm's performance becomes susceptible to entrapment in local minima.
    
    By systematically comparing the behavior of these iterative algorithms across multiple parameter regimes, our work reinforces the importance of algorithm-specific coefficient calibration and contributes to a more nuanced understanding of the optimization landscape in iterative ptychographic phase retrieval methods. The observation that performance degrades outside certain coefficient thresholds raises deeper questions about algorithm robustness, reproducibility, convergence dynamics, and susceptibility to local minimum areas that are not yet fully understood. Our contribution offers practical guidelines for effectively deploying these algorithms in experimental 4D-STEM applications, while also setting the stage for future studies aiming at improving reconstruction strategies. In summary, the insights presented here lay the groundwork for developing semi or fully automated iterative phase retrieval algorithms that are more robust and adaptable to changing conditions, such as varying dose levels.

\section*{CRediT authorship contribution}
    
    \textbf{T.C.}: Conceptualization, Software, Investigation, Formal analysis, Writing – original draft. \textbf{S.L.}: Software, Formal analysis, Investigation, review editing. \textbf{H.L.L.R.}: Conceptualization, Supervision, Writing – review editing. \textbf{C.H.}: Software, Formal analysis, Investigation, review. \textbf{N.J.S.}: Data acquisition, review. \textbf{S.B.}: Supervision, Writing – review editing. \textbf{T.J.P.}: Supervision, Conceptualization, Funding acquisition, Writing – review editing. \textbf{J.V.}: Funding acquisition, Supervision, Writing – review editing.

\section*{Declaration of competing interest}
    
    The authors declare that they have no known competing financial interests or personal relationships that could have appeared to influence the work reported in this paper.

\section*{Data availability}
    
    Data will be made available on request.

\section*{Acknowledgments}
    
    The authors kindly thank Prof. Andrew Maiden (EEE Department, University of Sheffield, Sheffield, United Kingdom) for assisting with the preliminary implementations of the original software and for his valuable insights.\\
    
    \textbf{J.V.} and \textbf{T.C.} acknowledge funding from the Flemish Government iBOF project PERsist. \textbf{S.B.} and \textbf{L.M.} acknowledge funding from the Research Foundation - Flanders (FWO) through project funding (Grant No. G0A7723N). \textbf{N.J.S.} acknowledges financial support from the Research Foundation - Flanders (FWO) through a postdoctoral fellowship (Grant No. 12AAO25N). \textbf{H.L.L.R.} and \textbf{J.V.} acknowledge further funding from the Horizon 2020 research and innovation programme (European Union), under grant agreement No 101017720 (FET-Proactive EBEAM). \textbf{J.V.} and \textbf{T.P.} acknowledge funding from the Flemish Government FWO project under grant No. G013122N.
    
\section*{Appendix A. Supplementary data}
Supplementary material related to this article can be found online.

\bibliographystyle{elsarticle-num}
\bibliography{bibliography}

\begin{thebibliography}{10}
\expandafter\ifx\csname url\endcsname\relax
  \def\url#1{\texttt{#1}}\fi
\expandafter\ifx\csname urlprefix\endcsname\relax\def\urlprefix{URL }\fi
\expandafter\ifx\csname href\endcsname\relax
  \def\href#1#2{#2} \def\path#1{#1}\fi

\bibitem{egerton_dose_2021}
R.~F. Egerton, \href{https://www.sciencedirect.com/science/article/pii/S0304399121001455}{Dose measurement in the {TEM} and {STEM}}, Ultramicroscopy 229 (2021) 113363.
\newblock \href {http://dx.doi.org/10.1016/j.ultramic.2021.113363} {\path{doi:10.1016/j.ultramic.2021.113363}}.
\newline\urlprefix\url{https://www.sciencedirect.com/science/article/pii/S0304399121001455}

\bibitem{erni_atomic-resolution_2009}
R.~Erni, M.~D. Rossell, C.~Kisielowski, U.~Dahmen, \href{https://link.aps.org/doi/10.1103/PhysRevLett.102.096101}{Atomic-{Resolution} {Imaging} with a {Sub}-50-pm {Electron} {Probe}}, Physical Review Letters 102~(9) (2009) 096101, publisher: American Physical Society.
\newblock \href {http://dx.doi.org/10.1103/PhysRevLett.102.096101} {\path{doi:10.1103/PhysRevLett.102.096101}}.
\newline\urlprefix\url{https://link.aps.org/doi/10.1103/PhysRevLett.102.096101}

\bibitem{batson_sub-angstrom_2002}
P.~E. Batson, N.~Dellby, O.~L. Krivanek, \href{https://www.nature.com/articles/nature00972}{Sub-ångstrom resolution using aberration corrected electron optics}, Nature 418~(6898) (2002) 617--620, publisher: Nature Publishing Group.
\newblock \href {http://dx.doi.org/10.1038/nature00972} {\path{doi:10.1038/nature00972}}.
\newline\urlprefix\url{https://www.nature.com/articles/nature00972}

\bibitem{chen_imaging_2020}
Q.~Chen, C.~Dwyer, G.~Sheng, C.~Zhu, X.~Li, C.~Zheng, Y.~Zhu, \href{https://onlinelibrary.wiley.com/doi/10.1002/adma.201907619}{Imaging {Beam}‐{Sensitive} {Materials} by {Electron} {Microscopy}}, Advanced Materials 32~(16) (2020) 1907619.
\newblock \href {http://dx.doi.org/10.1002/adma.201907619} {\path{doi:10.1002/adma.201907619}}.
\newline\urlprefix\url{https://onlinelibrary.wiley.com/doi/10.1002/adma.201907619}

\bibitem{egerton_mechanisms_2012}
R.~Egerton, \href{https://analyticalsciencejournals.onlinelibrary.wiley.com/doi/10.1002/jemt.22099}{Mechanisms of radiation damage in beam‐sensitive specimens, for {TEM} accelerating voltages between 10 and 300 {kV}}, Microscopy Research and Technique 75~(11) (2012) 1550--1556.
\newblock \href {http://dx.doi.org/10.1002/jemt.22099} {\path{doi:10.1002/jemt.22099}}.
\newline\urlprefix\url{https://analyticalsciencejournals.onlinelibrary.wiley.com/doi/10.1002/jemt.22099}

\bibitem{egerton_control_2013}
R.~Egerton, \href{https://linkinghub.elsevier.com/retrieve/pii/S0304399112001763}{Control of radiation damage in the {TEM}}, Ultramicroscopy 127 (2013) 100--108.
\newblock \href {http://dx.doi.org/10.1016/j.ultramic.2012.07.006} {\path{doi:10.1016/j.ultramic.2012.07.006}}.
\newline\urlprefix\url{https://linkinghub.elsevier.com/retrieve/pii/S0304399112001763}

\bibitem{egerton_radiation_2019}
R.~Egerton, \href{https://linkinghub.elsevier.com/retrieve/pii/S0968432818304359}{Radiation damage to organic and inorganic specimens in the {TEM}}, Micron 119 (2019) 72--87.
\newblock \href {http://dx.doi.org/10.1016/j.micron.2019.01.005} {\path{doi:10.1016/j.micron.2019.01.005}}.
\newline\urlprefix\url{https://linkinghub.elsevier.com/retrieve/pii/S0968432818304359}

\bibitem{susi_quantifying_2019}
T.~Susi, J.~C. Meyer, J.~Kotakoski, \href{https://www.nature.com/articles/s42254-019-0058-y}{Quantifying transmission electron microscopy irradiation effects using two-dimensional materials}, Nature Reviews Physics 1~(6) (2019) 397--405.
\newblock \href {http://dx.doi.org/10.1038/s42254-019-0058-y} {\path{doi:10.1038/s42254-019-0058-y}}.
\newline\urlprefix\url{https://www.nature.com/articles/s42254-019-0058-y}

\bibitem{meyer_accurate_2012}
J.~C. Meyer, F.~Eder, S.~Kurasch, V.~Skakalova, J.~Kotakoski, H.~J. Park, S.~Roth, A.~Chuvilin, S.~Eyhusen, G.~Benner, A.~V. Krasheninnikov, U.~Kaiser, \href{https://link.aps.org/doi/10.1103/PhysRevLett.108.196102}{Accurate {Measurement} of {Electron} {Beam} {Induced} {Displacement} {Cross} {Sections} for {Single}-{Layer} {Graphene}}, Physical Review Letters 108~(19) (2012) 196102.
\newblock \href {http://dx.doi.org/10.1103/PhysRevLett.108.196102} {\path{doi:10.1103/PhysRevLett.108.196102}}.
\newline\urlprefix\url{https://link.aps.org/doi/10.1103/PhysRevLett.108.196102}

\bibitem{scheid_electron_2023}
A.~Scheid, Y.~Wang, M.~Jung, T.~Heil, D.~Moia, J.~Maier, P.~A. Van~Aken, \href{https://academic.oup.com/mam/article/29/3/869/7131445}{Electron {Ptychographic} {Phase} {Imaging} of {Beam}-sensitive {All}-inorganic {Halide} {Perovskites} {Using} {Four}-dimensional {Scanning} {Transmission} {Electron} {Microscopy}}, Microscopy and Microanalysis 29~(3) (2023) 869--878.
\newblock \href {http://dx.doi.org/10.1093/micmic/ozad017} {\path{doi:10.1093/micmic/ozad017}}.
\newline\urlprefix\url{https://academic.oup.com/mam/article/29/3/869/7131445}

\bibitem{satta_formation_2021}
J.~Satta, A.~Casu, D.~Chiriu, C.~M. Carbonaro, L.~Stagi, P.~C. Ricci, \href{https://www.mdpi.com/2079-4991/11/7/1823}{Formation {Mechanisms} and {Phase} {Stability} of {Solid}-{State} {Grown} {CsPbI3} {Perovskites}}, Nanomaterials 11~(7) (2021) 1823, number: 7 Publisher: Multidisciplinary Digital Publishing Institute.
\newblock \href {http://dx.doi.org/10.3390/nano11071823} {\path{doi:10.3390/nano11071823}}.
\newline\urlprefix\url{https://www.mdpi.com/2079-4991/11/7/1823}

\bibitem{dhivyaprasath_degradation_2023}
K.~Dhivyaprasath, M.~Ashok, \href{https://linkinghub.elsevier.com/retrieve/pii/S0038092X23001767}{Degradation {Behavior} of {Methylammonium} {Lead} {Iodide} ({CH3NH3PbI3}) {Perovskite} {Film} in {Ambient} {Atmosphere} and {Device}}, Solar Energy 255 (2023) 89--98.
\newblock \href {http://dx.doi.org/10.1016/j.solener.2023.03.021} {\path{doi:10.1016/j.solener.2023.03.021}}.
\newline\urlprefix\url{https://linkinghub.elsevier.com/retrieve/pii/S0038092X23001767}

\bibitem{schrenker_investigation_2024}
N.~J. Schrenker, T.~Braeckevelt, A.~De~Backer, N.~Livakas, C.-P. Yu, T.~Friedrich, M.~B.~J. Roeffaers, J.~Hofkens, J.~Verbeeck, L.~Manna, V.~Van~Speybroeck, S.~Van~Aert, S.~Bals, \href{https://pubs.acs.org/doi/10.1021/acs.nanolett.4c02811}{Investigation of the {Octahedral} {Network} {Structure} in {Formamidinium} {Lead} {Bromide} {Nanocrystals} by {Low}-{Dose} {Scanning} {Transmission} {Electron} {Microscopy}}, Nano Letters 24~(35) (2024) 10936--10942.
\newblock \href {http://dx.doi.org/10.1021/acs.nanolett.4c02811} {\path{doi:10.1021/acs.nanolett.4c02811}}.
\newline\urlprefix\url{https://pubs.acs.org/doi/10.1021/acs.nanolett.4c02811}

\bibitem{ugurlu_radiolysis_2011}
O.~Ugurlu, J.~Haus, A.~A. Gunawan, M.~G. Thomas, S.~Maheshwari, M.~Tsapatsis, K.~A. Mkhoyan, \href{https://link.aps.org/doi/10.1103/PhysRevB.83.113408}{Radiolysis to knock-on damage transition in zeolites under electron beam irradiation}, Physical Review B 83~(11) (2011) 113408.
\newblock \href {http://dx.doi.org/10.1103/PhysRevB.83.113408} {\path{doi:10.1103/PhysRevB.83.113408}}.
\newline\urlprefix\url{https://link.aps.org/doi/10.1103/PhysRevB.83.113408}

\bibitem{liu_direct_2020}
L.~Liu, N.~Wang, C.~Zhu, X.~Liu, Y.~Zhu, P.~Guo, L.~Alfilfil, X.~Dong, D.~Zhang, Y.~Han, \href{https://onlinelibrary.wiley.com/doi/abs/10.1002/anie.201909834}{Direct {Imaging} of {Atomically} {Dispersed} {Molybdenum} that {Enables} {Location} of {Aluminum} in the {Framework} of {Zeolite} {ZSM}-5}, Angewandte Chemie International Edition 59~(2) (2020) 819--825, \_eprint: https://onlinelibrary.wiley.com/doi/pdf/10.1002/anie.201909834.
\newblock \href {http://dx.doi.org/10.1002/anie.201909834} {\path{doi:10.1002/anie.201909834}}.
\newline\urlprefix\url{https://onlinelibrary.wiley.com/doi/abs/10.1002/anie.201909834}

\bibitem{ooe_direct_2023}
K.~Ooe, T.~Seki, K.~Yoshida, Y.~Kohno, Y.~Ikuhara, N.~Shibata, \href{https://www.science.org/doi/full/10.1126/sciadv.adf6865}{Direct imaging of local atomic structures in zeolite using optimum bright-field scanning transmission electron microscopy}, Science Advances 9~(31) (2023) eadf6865, publisher: American Association for the Advancement of Science.
\newblock \href {http://dx.doi.org/10.1126/sciadv.adf6865} {\path{doi:10.1126/sciadv.adf6865}}.
\newline\urlprefix\url{https://www.science.org/doi/full/10.1126/sciadv.adf6865}

\bibitem{furukawa_chemistry_2013}
H.~Furukawa, K.~E. Cordova, M.~O’Keeffe, O.~M. Yaghi, \href{https://www.science.org/doi/10.1126/science.1230444}{The {Chemistry} and {Applications} of {Metal}-{Organic} {Frameworks}}, Science 341~(6149) (2013) 1230444, publisher: American Association for the Advancement of Science.
\newblock \href {http://dx.doi.org/10.1126/science.1230444} {\path{doi:10.1126/science.1230444}}.
\newline\urlprefix\url{https://www.science.org/doi/10.1126/science.1230444}

\bibitem{yang_catalysis_2019}
D.~Yang, B.~C. Gates, \href{https://doi.org/10.1021/acscatal.8b04515}{Catalysis by {Metal} {Organic} {Frameworks}: {Perspective} and {Suggestions} for {Future} {Research}}, ACS Catalysis 9~(3) (2019) 1779--1798, publisher: American Chemical Society.
\newblock \href {http://dx.doi.org/10.1021/acscatal.8b04515} {\path{doi:10.1021/acscatal.8b04515}}.
\newline\urlprefix\url{https://doi.org/10.1021/acscatal.8b04515}

\bibitem{bavykina_metalorganic_2020}
A.~Bavykina, N.~Kolobov, I.~S. Khan, J.~A. Bau, A.~Ramirez, J.~Gascon, \href{https://doi.org/10.1021/acs.chemrev.9b00685}{Metal–{Organic} {Frameworks} in {Heterogeneous} {Catalysis}: {Recent} {Progress}, {New} {Trends}, and {Future} {Perspectives}}, Chemical Reviews 120~(16) (2020) 8468--8535, publisher: American Chemical Society.
\newblock \href {http://dx.doi.org/10.1021/acs.chemrev.9b00685} {\path{doi:10.1021/acs.chemrev.9b00685}}.
\newline\urlprefix\url{https://doi.org/10.1021/acs.chemrev.9b00685}

\bibitem{kavak_high-resolution_2025}
S.~Kavak, D.~Jannis, A.~De~Backer, D.~A. Esteban, A.~Annys, S.~Carrasco, J.~Ferrando-Ferrero, R.~M. Guerrero, P.~Horcajada, J.~Verbeeck, S.~Van~Aert, S.~Bals, \href{https://xlink.rsc.org/?DOI=D4TA06724J}{High-resolution electron microscopy imaging of {MOFs} at optimized electron dose}, Journal of Materials Chemistry A (2025) 10.1039.D4TA06724J\href {http://dx.doi.org/10.1039/D4TA06724J} {\path{doi:10.1039/D4TA06724J}}.
\newline\urlprefix\url{https://xlink.rsc.org/?DOI=D4TA06724J}

\bibitem{glaeser_limitations_1971}
R.~M. Glaeser, \href{https://www.sciencedirect.com/science/article/pii/S0022532071801181}{Limitations to significant information in biological electron microscopy as a result of radiation damage}, Journal of Ultrastructure Research 36~(3) (1971) 466--482.
\newblock \href {http://dx.doi.org/10.1016/S0022-5320(71)80118-1} {\path{doi:10.1016/S0022-5320(71)80118-1}}.
\newline\urlprefix\url{https://www.sciencedirect.com/science/article/pii/S0022532071801181}

\bibitem{henderson_three-dimensional_1975}
R.~Henderson, P.~N.~T. Unwin, \href{https://www.nature.com/articles/257028a0}{Three-dimensional model of purple membrane obtained by electron microscopy}, Nature 257~(5521) (1975) 28--32, publisher: Nature Publishing Group.
\newblock \href {http://dx.doi.org/10.1038/257028a0} {\path{doi:10.1038/257028a0}}.
\newline\urlprefix\url{https://www.nature.com/articles/257028a0}

\bibitem{kucukoglu_low-dose_2024}
B.~Küçükoğlu, I.~Mohammed, R.~C. Guerrero-Ferreira, S.~M. Ribet, G.~Varnavides, M.~L. Leidl, K.~Lau, S.~Nazarov, A.~Myasnikov, M.~Kube, J.~Radecke, C.~Sachse, K.~Müller-Caspary, C.~Ophus, H.~Stahlberg, \href{https://www.nature.com/articles/s41467-024-52403-5}{Low-dose cryo-electron ptychography of proteins at sub-nanometer resolution}, Nature Communications 15~(1) (2024) 8062, publisher: Nature Publishing Group.
\newblock \href {http://dx.doi.org/10.1038/s41467-024-52403-5} {\path{doi:10.1038/s41467-024-52403-5}}.
\newline\urlprefix\url{https://www.nature.com/articles/s41467-024-52403-5}

\bibitem{zhou_low-dose_2020}
L.~Zhou, J.~Song, J.~S. Kim, X.~Pei, C.~Huang, M.~Boyce, L.~Mendonça, D.~Clare, A.~Siebert, C.~S. Allen, E.~Liberti, D.~Stuart, X.~Pan, P.~D. Nellist, P.~Zhang, A.~I. Kirkland, P.~Wang, \href{https://www.nature.com/articles/s41467-020-16391-6}{Low-dose phase retrieval of biological specimens using cryo-electron ptychography}, Nature Communications 11~(1) (2020) 2773, publisher: Nature Publishing Group.
\newblock \href {http://dx.doi.org/10.1038/s41467-020-16391-6} {\path{doi:10.1038/s41467-020-16391-6}}.
\newline\urlprefix\url{https://www.nature.com/articles/s41467-020-16391-6}

\bibitem{lazic_phase_2016}
I.~Lazić, E.~G.~T. Bosch, S.~Lazar, \href{https://www.sciencedirect.com/science/article/pii/S0304399115300449}{Phase contrast {STEM} for thin samples: {Integrated} differential phase contrast}, Ultramicroscopy 160 (2016) 265--280.
\newblock \href {http://dx.doi.org/10.1016/j.ultramic.2015.10.011} {\path{doi:10.1016/j.ultramic.2015.10.011}}.
\newline\urlprefix\url{https://www.sciencedirect.com/science/article/pii/S0304399115300449}

\bibitem{lazic_single-particle_2022}
I.~Lazić, M.~Wirix, M.~L. Leidl, F.~De~Haas, D.~Mann, M.~Beckers, E.~V. Pechnikova, K.~Müller-Caspary, R.~Egoavil, E.~G.~T. Bosch, C.~Sachse, \href{https://www.nature.com/articles/s41592-022-01586-0}{Single-particle cryo-{EM} structures from {iDPC}–{STEM} at near-atomic resolution}, Nature Methods 19~(9) (2022) 1126--1136.
\newblock \href {http://dx.doi.org/10.1038/s41592-022-01586-0} {\path{doi:10.1038/s41592-022-01586-0}}.
\newline\urlprefix\url{https://www.nature.com/articles/s41592-022-01586-0}

\bibitem{li_direct_2019}
X.~Li, J.~Wang, X.~Liu, L.~Liu, D.~Cha, X.~Zheng, A.~A. Yousef, K.~Song, Y.~Zhu, D.~Zhang, Y.~Han, \href{https://doi.org/10.1021/jacs.9b04896}{Direct {Imaging} of {Tunable} {Crystal} {Surface} {Structures} of {MOF} {MIL}-101 {Using} {High}-{Resolution} {Electron} {Microscopy}}, Journal of the American Chemical Society 141~(30) (2019) 12021--12028, publisher: American Chemical Society.
\newblock \href {http://dx.doi.org/10.1021/jacs.9b04896} {\path{doi:10.1021/jacs.9b04896}}.
\newline\urlprefix\url{https://doi.org/10.1021/jacs.9b04896}

\bibitem{yang_efficient_2015-1}
H.~Yang, T.~J. Pennycook, P.~D. Nellist, \href{https://linkinghub.elsevier.com/retrieve/pii/S0304399114002058}{Efficient phase contrast imaging in {STEM} using a pixelated detector. {Part} {II}: {Optimisation} of imaging conditions}, Ultramicroscopy 151 (2015) 232--239.
\newblock \href {http://dx.doi.org/10.1016/j.ultramic.2014.10.013} {\path{doi:10.1016/j.ultramic.2014.10.013}}.
\newline\urlprefix\url{https://linkinghub.elsevier.com/retrieve/pii/S0304399114002058}

\bibitem{oleary_phase_2020}
C.~M. O'Leary, C.~S. Allen, C.~Huang, J.~S. Kim, E.~Liberti, P.~D. Nellist, A.~I. Kirkland, \href{https://doi.org/10.1063/1.5143213}{Phase reconstruction using fast binary {4D} {STEM} data}, Applied Physics Letters 116~(12) (2020) 124101.
\newblock \href {http://dx.doi.org/10.1063/1.5143213} {\path{doi:10.1063/1.5143213}}.
\newline\urlprefix\url{https://doi.org/10.1063/1.5143213}

\bibitem{oleary_contrast_2021}
C.~M. O’Leary, G.~T. Martinez, E.~Liberti, M.~J. Humphry, A.~I. Kirkland, P.~D. Nellist, \href{https://www.sciencedirect.com/science/article/pii/S0304399120303314}{Contrast transfer and noise considerations in focused-probe electron ptychography}, Ultramicroscopy 221 (2021) 113189.
\newblock \href {http://dx.doi.org/10.1016/j.ultramic.2020.113189} {\path{doi:10.1016/j.ultramic.2020.113189}}.
\newline\urlprefix\url{https://www.sciencedirect.com/science/article/pii/S0304399120303314}

\bibitem{hao_atomic-scale_2023}
B.~Hao, Z.~Ding, X.~Tao, P.~D. Nellist, H.~E. Assender, \href{https://www.sciencedirect.com/science/article/pii/S0032386123006353}{Atomic-scale imaging of polyvinyl alcohol crystallinity using electron ptychography}, Polymer 284 (2023) 126305.
\newblock \href {http://dx.doi.org/10.1016/j.polymer.2023.126305} {\path{doi:10.1016/j.polymer.2023.126305}}.
\newline\urlprefix\url{https://www.sciencedirect.com/science/article/pii/S0032386123006353}

\bibitem{dong_atomic-level_2023}
Z.~Dong, E.~Zhang, Y.~Jiang, Q.~Zhang, A.~Mayoral, H.~Jiang, Y.~Ma, \href{https://doi.org/10.1021/jacs.2c12673}{Atomic-{Level} {Imaging} of {Zeolite} {Local} {Structures} {Using} {Electron} {Ptychography}}, Journal of the American Chemical Society 145~(12) (2023) 6628--6632, publisher: American Chemical Society.
\newblock \href {http://dx.doi.org/10.1021/jacs.2c12673} {\path{doi:10.1021/jacs.2c12673}}.
\newline\urlprefix\url{https://doi.org/10.1021/jacs.2c12673}

\bibitem{robert_benchmarking_2025}
H.~L.~L. Robert, M.~L. Leidl, K.~Müller-Caspary, J.~Verbeeck, \href{http://arxiv.org/abs/2501.08874}{Benchmarking analytical electron ptychography methods for the low-dose imaging of beam-sensitive materials}, arXiv:2501.08874 [physics] (Jan. 2025).
\newblock \href {http://dx.doi.org/10.48550/arXiv.2501.08874} {\path{doi:10.48550/arXiv.2501.08874}}.
\newline\urlprefix\url{http://arxiv.org/abs/2501.08874}

\bibitem{li_atomically_2025}
G.~Li, M.~Xu, W.-Q. Tang, Y.~Liu, C.~Chen, D.~Zhang, L.~Liu, S.~Ning, H.~Zhang, Z.-Y. Gu, Z.~Lai, D.~A. Muller, Y.~Han, \href{https://www.nature.com/articles/s41467-025-56215-z}{Atomically resolved imaging of radiation-sensitive metal-organic frameworks via electron ptychography}, Nature Communications 16~(1) (2025) 914, publisher: Nature Publishing Group.
\newblock \href {http://dx.doi.org/10.1038/s41467-025-56215-z} {\path{doi:10.1038/s41467-025-56215-z}}.
\newline\urlprefix\url{https://www.nature.com/articles/s41467-025-56215-z}

\bibitem{lozano_low-dose_2018}
J.~G. Lozano, G.~T. Martinez, L.~Jin, P.~D. Nellist, P.~G. Bruce, \href{https://doi.org/10.1021/acs.nanolett.8b02718}{Low-{Dose} {Aberration}-{Free} {Imaging} of {Li}-{Rich} {Cathode} {Materials} at {Various} {States} of {Charge} {Using} {Electron} {Ptychography}}, Nano Letters 18~(11) (2018) 6850--6855, publisher: American Chemical Society.
\newblock \href {http://dx.doi.org/10.1021/acs.nanolett.8b02718} {\path{doi:10.1021/acs.nanolett.8b02718}}.
\newline\urlprefix\url{https://doi.org/10.1021/acs.nanolett.8b02718}

\bibitem{D'Alfonso_deterministic}
A.~J. D'Alfonso, A.~J. Morgan, A.~W.~C. Yan, P.~Wang, H.~Sawada, A.~I. Kirkland, L.~J. Allen, Deterministic electron ptychography at atomic resolution, PHYSICAL REVIEW B 89~(6).
\newblock \href {http://dx.doi.org/10.1103/PhysRevB.89.064101} {\path{doi:10.1103/PhysRevB.89.064101}}.

\bibitem{D'Alfonso_dose_dependent}
A.~J. D'Alfonso, L.~J. Allen, H.~Sawada, A.~I. Kirkland, Dose-dependent high-resolution electron ptychography, JOURNAL OF APPLIED PHYSICS 119~(5).
\newblock \href {http://dx.doi.org/10.1063/1.4941269} {\path{doi:10.1063/1.4941269}}.

\bibitem{fienup_reconstruction_1978}
J.~R. Fienup, \href{https://opg.optica.org/abstract.cfm?URI=ol-3-1-27}{Reconstruction of an object from the modulus of its {Fourier} transform}, Optics Letters 3~(1) (1978) 27.
\newblock \href {http://dx.doi.org/10.1364/OL.3.000027} {\path{doi:10.1364/OL.3.000027}}.
\newline\urlprefix\url{https://opg.optica.org/abstract.cfm?URI=ol-3-1-27}

\bibitem{fienup_phase_1982}
J.~R. Fienup, \href{https://opg.optica.org/abstract.cfm?URI=ao-21-15-2758}{Phase retrieval algorithms: a comparison}, Applied Optics 21~(15) (1982) 2758.
\newblock \href {http://dx.doi.org/10.1364/AO.21.002758} {\path{doi:10.1364/AO.21.002758}}.
\newline\urlprefix\url{https://opg.optica.org/abstract.cfm?URI=ao-21-15-2758}

\bibitem{fienup_reconstruction_1987}
J.~R. Fienup, \href{https://opg.optica.org/abstract.cfm?URI=josaa-4-1-118}{Reconstruction of a complex-valued object from the modulus of its {Fourier} transform using a support constraint}, Journal of the Optical Society of America A 4~(1) (1987) 118.
\newblock \href {http://dx.doi.org/10.1364/JOSAA.4.000118} {\path{doi:10.1364/JOSAA.4.000118}}.
\newline\urlprefix\url{https://opg.optica.org/abstract.cfm?URI=josaa-4-1-118}

\bibitem{miao_phase_1998}
J.~Miao, D.~Sayre, H.~N. Chapman, \href{https://opg.optica.org/abstract.cfm?URI=josaa-15-6-1662}{Phase retrieval from the magnitude of the {Fourier} transforms of nonperiodic objects}, Journal of the Optical Society of America A 15~(6) (1998) 1662.
\newblock \href {http://dx.doi.org/10.1364/JOSAA.15.001662} {\path{doi:10.1364/JOSAA.15.001662}}.
\newline\urlprefix\url{https://opg.optica.org/abstract.cfm?URI=josaa-15-6-1662}

\bibitem{miao_extending_1999}
J.~Miao, P.~Charalambous, J.~Kirz, D.~Sayre, \href{https://www.nature.com/articles/22498}{Extending the methodology of {X}-ray crystallography to allow imaging of micrometre-sized non-crystalline specimens}, Nature 400~(6742) (1999) 342--344.
\newblock \href {http://dx.doi.org/10.1038/22498} {\path{doi:10.1038/22498}}.
\newline\urlprefix\url{https://www.nature.com/articles/22498}

\bibitem{hoppe_beugung_1969}
W.~Hoppe, \href{https://scripts.iucr.org/cgi-bin/paper?S0567739469001069}{Beugung im inhomogenen {Primärstrahlwellenfeld}. {III}. {Amplituden}- und {Phasenbestimmung} bei unperiodischen {Objekten}}, Acta Crystallographica Section A 25~(4) (1969) 508--514.
\newblock \href {http://dx.doi.org/10.1107/S0567739469001069} {\path{doi:10.1107/S0567739469001069}}.
\newline\urlprefix\url{https://scripts.iucr.org/cgi-bin/paper?S0567739469001069}

\bibitem{hoppe_beugung_1969-1}
W.~Hoppe, G.~Strube, \href{https://scripts.iucr.org/cgi-bin/paper?S0567739469001057}{Beugung in inhomogenen {Primärstrahlenwellenfeld}. {II}. {Lichtoptische} {Analogieversuche} zur {Phasenmessung} von {Gitterinterferenzen}}, Acta Crystallographica Section A 25~(4) (1969) 502--507.
\newblock \href {http://dx.doi.org/10.1107/S0567739469001057} {\path{doi:10.1107/S0567739469001057}}.
\newline\urlprefix\url{https://scripts.iucr.org/cgi-bin/paper?S0567739469001057}

\bibitem{hoppe_beugung_1969-2}
W.~Hoppe, \href{https://scripts.iucr.org/cgi-bin/paper?S0567739469001045}{Beugung im inhomogenen {Primärstrahlwellenfeld}. {I}. {Prinzip} einer {Phasenmessung} von {Elektronenbeungungsinterferenzen}}, Acta Crystallographica Section A 25~(4) (1969) 495--501.
\newblock \href {http://dx.doi.org/10.1107/S0567739469001045} {\path{doi:10.1107/S0567739469001045}}.
\newline\urlprefix\url{https://scripts.iucr.org/cgi-bin/paper?S0567739469001045}

\bibitem{mcmullan_electron_2007}
G.~McMullan, D.~M. Cattermole, S.~Chen, R.~Henderson, X.~Llopart, C.~Summerfield, L.~Tlustos, A.~R. Faruqi, \href{https://www.sciencedirect.com/science/article/pii/S0304399106001963}{Electron imaging with {Medipix2} hybrid pixel detector}, Ultramicroscopy 107~(4) (2007) 401--413.
\newblock \href {http://dx.doi.org/10.1016/j.ultramic.2006.10.005} {\path{doi:10.1016/j.ultramic.2006.10.005}}.
\newline\urlprefix\url{https://www.sciencedirect.com/science/article/pii/S0304399106001963}

\bibitem{llopart_timepix_2007}
X.~Llopart, R.~Ballabriga, M.~Campbell, L.~Tlustos, W.~Wong, \href{https://www.sciencedirect.com/science/article/pii/S0168900207017020}{Timepix, a 65k programmable pixel readout chip for arrival time, energy and/or photon counting measurements}, Nuclear Instruments and Methods in Physics Research Section A: Accelerators, Spectrometers, Detectors and Associated Equipment 581~(1) (2007) 485--494.
\newblock \href {http://dx.doi.org/10.1016/j.nima.2007.08.079} {\path{doi:10.1016/j.nima.2007.08.079}}.
\newline\urlprefix\url{https://www.sciencedirect.com/science/article/pii/S0168900207017020}

\bibitem{ballabriga_medipix3_2011}
R.~Ballabriga, M.~Campbell, E.~Heijne, X.~Llopart, L.~Tlustos, W.~Wong, \href{https://www.sciencedirect.com/science/article/pii/S0168900210012982}{Medipix3: {A} 64   k pixel detector readout chip working in single photon counting mode with improved spectrometric performance}, Nuclear Instruments and Methods in Physics Research Section A: Accelerators, Spectrometers, Detectors and Associated Equipment 633 (2011) S15--S18.
\newblock \href {http://dx.doi.org/10.1016/j.nima.2010.06.108} {\path{doi:10.1016/j.nima.2010.06.108}}.
\newline\urlprefix\url{https://www.sciencedirect.com/science/article/pii/S0168900210012982}

\bibitem{mir_characterisation_2017}
J.~A. Mir, R.~Clough, R.~MacInnes, C.~Gough, R.~Plackett, I.~Shipsey, H.~Sawada, I.~MacLaren, R.~Ballabriga, D.~Maneuski, V.~O'Shea, D.~McGrouther, A.~I. Kirkland, \href{https://www.sciencedirect.com/science/article/pii/S0304399116303989}{Characterisation of the {Medipix3} detector for 60 and 80   {keV} electrons}, Ultramicroscopy 182 (2017) 44--53.
\newblock \href {http://dx.doi.org/10.1016/j.ultramic.2017.06.010} {\path{doi:10.1016/j.ultramic.2017.06.010}}.
\newline\urlprefix\url{https://www.sciencedirect.com/science/article/pii/S0304399116303989}

\bibitem{ryll_pnccd-based_2016}
H.~Ryll, M.~Simson, R.~Hartmann, P.~Holl, M.~Huth, S.~Ihle, Y.~Kondo, P.~Kotula, A.~Liebel, K.~Müller-Caspary, A.~Rosenauer, R.~Sagawa, J.~Schmidt, H.~Soltau, L.~Strüder, \href{https://dx.doi.org/10.1088/1748-0221/11/04/P04006}{A {pnCCD}-based, fast direct single electron imaging camera for {TEM} and {STEM}}, Journal of Instrumentation 11~(04) (2016) P04006.
\newblock \href {http://dx.doi.org/10.1088/1748-0221/11/04/P04006} {\path{doi:10.1088/1748-0221/11/04/P04006}}.
\newline\urlprefix\url{https://dx.doi.org/10.1088/1748-0221/11/04/P04006}

\bibitem{philipp_very-high_2022}
H.~T. Philipp, M.~W. Tate, K.~S. Shanks, L.~Mele, M.~Peemen, P.~Dona, R.~Hartong, G.~van Veen, Y.-T. Shao, Z.~Chen, J.~Thom-Levy, D.~A. Muller, S.~M. Gruner, \href{https://doi.org/10.1017/S1431927622000174}{Very-{High} {Dynamic} {Range}, 10,000 {Frames}/{Second} {Pixel} {Array} {Detector} for {Electron} {Microscopy}}, Microscopy and Microanalysis 28~(2) (2022) 425--440.
\newblock \href {http://dx.doi.org/10.1017/S1431927622000174} {\path{doi:10.1017/S1431927622000174}}.
\newline\urlprefix\url{https://doi.org/10.1017/S1431927622000174}

\bibitem{zambon_enhanced_2023}
P.~Zambon, \href{https://www.frontiersin.org/journals/physics/articles/10.3389/fphy.2023.1123787/full}{Enhanced {DQE} and sub-pixel resolution by single-event processing in counting hybrid pixel electron detectors: {A} simulation study}, Frontiers in Physics 11, publisher: Frontiers.
\newblock \href {http://dx.doi.org/10.3389/fphy.2023.1123787} {\path{doi:10.3389/fphy.2023.1123787}}.
\newline\urlprefix\url{https://www.frontiersin.org/journals/physics/articles/10.3389/fphy.2023.1123787/full}

\bibitem{ercius_4d_2024}
P.~Ercius, I.~J. Johnson, P.~Pelz, B.~H. Savitzky, L.~Hughes, H.~G. Brown, S.~E. Zeltmann, S.-L. Hsu, C.~C.~S. Pedroso, B.~E. Cohen, R.~Ramesh, D.~Paul, J.~M. Joseph, T.~Stezelberger, C.~Czarnik, M.~Lent, E.~Fong, J.~Ciston, M.~C. Scott, C.~Ophus, A.~M. Minor, P.~Denes, The {4D} {Camera}: {An} 87 {kHz} {Direct} {Electron} {Detector} for {Scanning}/{Transmission} {Electron} {Microscopy}, Microscopy and Microanalysis: The Official Journal of Microscopy Society of America, Microbeam Analysis Society, Microscopical Society of Canada 30~(5) (2024) 903--912.
\newblock \href {http://dx.doi.org/10.1093/mam/ozae086} {\path{doi:10.1093/mam/ozae086}}.

\bibitem{yang_4d_2015}
H.~Yang, L.~Jones, H.~Ryll, M.~Simson, H.~Soltau, Y.~Kondo, R.~Sagawa, H.~Banba, I.~MacLaren, P.~D. Nellist, \href{https://iopscience.iop.org/article/10.1088/1742-6596/644/1/012032}{{4D} {STEM}: {High} efficiency phase contrast imaging using a fast pixelated detector}, Journal of Physics: Conference Series 644 (2015) 012032.
\newblock \href {http://dx.doi.org/10.1088/1742-6596/644/1/012032} {\path{doi:10.1088/1742-6596/644/1/012032}}.
\newline\urlprefix\url{https://iopscience.iop.org/article/10.1088/1742-6596/644/1/012032}

\bibitem{poikela_timepix3_2014}
T.~Poikela, J.~Plosila, T.~Westerlund, M.~Campbell, M.~D. Gaspari, X.~Llopart, V.~Gromov, R.~Kluit, M.~v. Beuzekom, F.~Zappon, V.~Zivkovic, C.~Brezina, K.~Desch, Y.~Fu, A.~Kruth, \href{https://dx.doi.org/10.1088/1748-0221/9/05/C05013}{Timepix3: a {65K} channel hybrid pixel readout chip with simultaneous {ToA}/{ToT} and sparse readout}, Journal of Instrumentation 9~(05) (2014) C05013.
\newblock \href {http://dx.doi.org/10.1088/1748-0221/9/05/C05013} {\path{doi:10.1088/1748-0221/9/05/C05013}}.
\newline\urlprefix\url{https://dx.doi.org/10.1088/1748-0221/9/05/C05013}

\bibitem{llopart_timepix4_2022}
X.~Llopart, J.~Alozy, R.~Ballabriga, M.~Campbell, R.~Casanova, V.~Gromov, E.~H.~M. Heijne, T.~Poikela, E.~Santin, V.~Sriskaran, L.~Tlustos, A.~Vitkovskiy, \href{https://dx.doi.org/10.1088/1748-0221/17/01/C01044}{Timepix4, a large area pixel detector readout chip which can be tiled on 4 sides providing sub-200 ps timestamp binning}, Journal of Instrumentation 17~(01) (2022) C01044, publisher: IOP Publishing.
\newblock \href {http://dx.doi.org/10.1088/1748-0221/17/01/C01044} {\path{doi:10.1088/1748-0221/17/01/C01044}}.
\newline\urlprefix\url{https://dx.doi.org/10.1088/1748-0221/17/01/C01044}

\bibitem{jannis_event_2022}
D.~Jannis, C.~Hofer, C.~Gao, X.~Xie, A.~Béché, T.~J. Pennycook, J.~Verbeeck, \href{https://www.sciencedirect.com/science/article/pii/S0304399121001996}{Event driven {4D} {STEM} acquisition with a {Timepix3} detector: {Microsecond} dwell time and faster scans for high precision and low dose applications}, Ultramicroscopy 233 (2022) 113423.
\newblock \href {http://dx.doi.org/10.1016/j.ultramic.2021.113423} {\path{doi:10.1016/j.ultramic.2021.113423}}.
\newline\urlprefix\url{https://www.sciencedirect.com/science/article/pii/S0304399121001996}

\bibitem{auad_time_2024}
Y.~Auad, J.~Baaboura, J.-D. Blazit, M.~Tencé, O.~Stéphan, M.~Kociak, L.~H.~G. Tizei, \href{https://www.sciencedirect.com/science/article/pii/S0304399123002061}{Time calibration studies for the {Timepix3} hybrid pixel detector in electron microscopy}, Ultramicroscopy 257 (2024) 113889.
\newblock \href {http://dx.doi.org/10.1016/j.ultramic.2023.113889} {\path{doi:10.1016/j.ultramic.2023.113889}}.
\newline\urlprefix\url{https://www.sciencedirect.com/science/article/pii/S0304399123002061}

\bibitem{denisov_characterization_2023}
N.~Denisov, D.~Jannis, A.~Orekhov, K.~Müller-Caspary, J.~Verbeeck, \href{https://www.sciencedirect.com/science/article/pii/S0304399123000943}{Characterization of a {Timepix} detector for use in {SEM} acceleration voltage range}, Ultramicroscopy 253 (2023) 113777.
\newblock \href {http://dx.doi.org/10.1016/j.ultramic.2023.113777} {\path{doi:10.1016/j.ultramic.2023.113777}}.
\newline\urlprefix\url{https://www.sciencedirect.com/science/article/pii/S0304399123000943}

\bibitem{rodenburg_theory_1992}
J.~M. Rodenburg, R.~H.~T. Bates, \href{https://royalsocietypublishing.org/doi/10.1098/rsta.1992.0050}{The theory of super-resolution electron microscopy via {Wigner}-distribution deconvolution}, Philosophical Transactions of the Royal Society of London. Series A: Physical and Engineering Sciences 339~(1655) (1992) 521--553, publisher: Royal Society.
\newblock \href {http://dx.doi.org/10.1098/rsta.1992.0050} {\path{doi:10.1098/rsta.1992.0050}}.
\newline\urlprefix\url{https://royalsocietypublishing.org/doi/10.1098/rsta.1992.0050}

\bibitem{bates_sub-angstrom_1989}
R.~H.~T. Bates, J.~M. Rodenburg, \href{https://www.sciencedirect.com/science/article/pii/0304399189900521}{Sub-ångström transmission microscopy: {A} fourier transform algorithm for microdiffraction plane intensity information}, Ultramicroscopy 31~(3) (1989) 303--307.
\newblock \href {http://dx.doi.org/10.1016/0304-3991(89)90052-1} {\path{doi:10.1016/0304-3991(89)90052-1}}.
\newline\urlprefix\url{https://www.sciencedirect.com/science/article/pii/0304399189900521}

\bibitem{mccallum_two-dimensional_1992}
B.~C. McCallum, J.~M. Rodenburg, \href{https://www.sciencedirect.com/science/article/pii/030439919290149E}{Two-dimensional demonstration of {Wigner} phase-retrieval microscopy in the {STEM} configuration}, Ultramicroscopy 45~(3) (1992) 371--380.
\newblock \href {http://dx.doi.org/10.1016/0304-3991(92)90149-E} {\path{doi:10.1016/0304-3991(92)90149-E}}.
\newline\urlprefix\url{https://www.sciencedirect.com/science/article/pii/030439919290149E}

\bibitem{rodenburg_experimental_1993}
J.~M. Rodenburg, B.~C. McCallum, P.~D. Nellist, \href{https://www.sciencedirect.com/science/article/pii/0304399193901057}{Experimental tests on double-resolution coherent imaging via {STEM}}, Ultramicroscopy 48~(3) (1993) 304--314.
\newblock \href {http://dx.doi.org/10.1016/0304-3991(93)90105-7} {\path{doi:10.1016/0304-3991(93)90105-7}}.
\newline\urlprefix\url{https://www.sciencedirect.com/science/article/pii/0304399193901057}

\bibitem{pennycook_efficient_2015}
T.~J. Pennycook, A.~R. Lupini, H.~Yang, M.~F. Murfitt, L.~Jones, P.~D. Nellist, \href{https://www.sciencedirect.com/science/article/pii/S0304399114001934}{Efficient phase contrast imaging in {STEM} using a pixelated detector. {Part} 1: {Experimental} demonstration at atomic resolution}, Ultramicroscopy 151 (2015) 160--167.
\newblock \href {http://dx.doi.org/10.1016/j.ultramic.2014.09.013} {\path{doi:10.1016/j.ultramic.2014.09.013}}.
\newline\urlprefix\url{https://www.sciencedirect.com/science/article/pii/S0304399114001934}

\bibitem{faulkner_movable_2004}
H.~M.~L. Faulkner, J.~M. Rodenburg, \href{https://link.aps.org/doi/10.1103/PhysRevLett.93.023903}{Movable {Aperture} {Lensless} {Transmission} {Microscopy}: {A} {Novel} {Phase} {Retrieval} {Algorithm}}, Physical Review Letters 93~(2) (2004) 023903.
\newblock \href {http://dx.doi.org/10.1103/PhysRevLett.93.023903} {\path{doi:10.1103/PhysRevLett.93.023903}}.
\newline\urlprefix\url{https://link.aps.org/doi/10.1103/PhysRevLett.93.023903}

\bibitem{rodenburg_phase_2004}
J.~M. Rodenburg, H.~M.~L. Faulkner, \href{https://pubs.aip.org/apl/article/85/20/4795/329473/A-phase-retrieval-algorithm-for-shifting}{A phase retrieval algorithm for shifting illumination}, Applied Physics Letters 85~(20) (2004) 4795--4797.
\newblock \href {http://dx.doi.org/10.1063/1.1823034} {\path{doi:10.1063/1.1823034}}.
\newline\urlprefix\url{https://pubs.aip.org/apl/article/85/20/4795/329473/A-phase-retrieval-algorithm-for-shifting}

\bibitem{maiden_improved_2009}
A.~M. Maiden, J.~M. Rodenburg, \href{https://linkinghub.elsevier.com/retrieve/pii/S0304399109001284}{An improved ptychographical phase retrieval algorithm for diffractive imaging}, Ultramicroscopy 109~(10) (2009) 1256--1262.
\newblock \href {http://dx.doi.org/10.1016/j.ultramic.2009.05.012} {\path{doi:10.1016/j.ultramic.2009.05.012}}.
\newline\urlprefix\url{https://linkinghub.elsevier.com/retrieve/pii/S0304399109001284}

\bibitem{maiden_further_2017}
A.~Maiden, D.~Johnson, P.~Li, \href{https://opg.optica.org/optica/abstract.cfm?uri=optica-4-7-736}{Further improvements to the ptychographical iterative engine}, Optica 4~(7) (2017) 736--745, publisher: Optica Publishing Group.
\newblock \href {http://dx.doi.org/10.1364/OPTICA.4.000736} {\path{doi:10.1364/OPTICA.4.000736}}.
\newline\urlprefix\url{https://opg.optica.org/optica/abstract.cfm?uri=optica-4-7-736}

\bibitem{thibault_probe_2009}
P.~Thibault, M.~Dierolf, O.~Bunk, A.~Menzel, F.~Pfeiffer, \href{https://www.sciencedirect.com/science/article/pii/S0304399108003458}{Probe retrieval in ptychographic coherent diffractive imaging}, Ultramicroscopy 109~(4) (2009) 338--343.
\newblock \href {http://dx.doi.org/10.1016/j.ultramic.2008.12.011} {\path{doi:10.1016/j.ultramic.2008.12.011}}.
\newline\urlprefix\url{https://www.sciencedirect.com/science/article/pii/S0304399108003458}

\bibitem{thibault_high-resolution_2008}
P.~Thibault, M.~Dierolf, A.~Menzel, O.~Bunk, C.~David, F.~Pfeiffer, \href{https://www.science.org/doi/10.1126/science.1158573}{High-{Resolution} {Scanning} {X}-ray {Diffraction} {Microscopy}}, Science 321~(5887) (2008) 379--382, publisher: American Association for the Advancement of Science.
\newblock \href {http://dx.doi.org/10.1126/science.1158573} {\path{doi:10.1126/science.1158573}}.
\newline\urlprefix\url{https://www.science.org/doi/10.1126/science.1158573}

\bibitem{maiden_wasp_2024}
A.~Maiden, W.~Mei, P.~Li, \href{https://preprints.opticaopen.org/articles/preprint/WASP_Weighted_Average_of_Sequential_Projections_for_ptychographic_phase_retrieval/24894489/1}{{WASP}: {Weighted} {Average} of {Sequential} {Projections} for ptychographic phase retrieval} (Jan. 2024).
\newblock \href {http://dx.doi.org/10.1364/opticaopen.24894489.v1} {\path{doi:10.1364/opticaopen.24894489.v1}}.
\newline\urlprefix\url{https://preprints.opticaopen.org/articles/preprint/WASP_Weighted_Average_of_Sequential_Projections_for_ptychographic_phase_retrieval/24894489/1}

\bibitem{cowley_electron_1972}
J.~M. Cowley, S.~Iijima, \href{https://www.degruyterbrill.com/document/doi/10.1515/zna-1972-0312/html}{Electron {Microscope} {Image} {Contrast} for {Thin} {Crystal}}, Zeitschrift für Naturforschung A 27~(3) (1972) 445--451, publisher: De Gruyter Section: Zeitschrift für Naturforschung A.
\newblock \href {http://dx.doi.org/10.1515/zna-1972-0312} {\path{doi:10.1515/zna-1972-0312}}.
\newline\urlprefix\url{https://www.degruyterbrill.com/document/doi/10.1515/zna-1972-0312/html}

\bibitem{luczka_master_1991}
J.~Luczka, M.~Niemiec, \href{https://dx.doi.org/10.1088/0305-4470/24/17/010}{A master equation for quantum systems driven by {Poisson} white noise}, Journal of Physics A: Mathematical and General 24~(17) (1991) L1021.
\newblock \href {http://dx.doi.org/10.1088/0305-4470/24/17/010} {\path{doi:10.1088/0305-4470/24/17/010}}.
\newline\urlprefix\url{https://dx.doi.org/10.1088/0305-4470/24/17/010}

\bibitem{seki_theoretical_2018}
T.~Seki, Y.~Ikuhara, N.~Shibata, \href{https://www.sciencedirect.com/science/article/pii/S0304399118300603}{Theoretical framework of statistical noise in scanning transmission electron microscopy}, Ultramicroscopy 193 (2018) 118--125.
\newblock \href {http://dx.doi.org/10.1016/j.ultramic.2018.06.014} {\path{doi:10.1016/j.ultramic.2018.06.014}}.
\newline\urlprefix\url{https://www.sciencedirect.com/science/article/pii/S0304399118300603}

\bibitem{weierstall_image_2002}
U.~Weierstall, Q.~Chen, J.~C.~H. Spence, M.~R. Howells, M.~Isaacson, R.~R. Panepucci, \href{https://www.sciencedirect.com/science/article/pii/S0304399101001346}{Image reconstruction from electron and {X}-ray diffraction patterns using iterative algorithms: experiment and simulation}, Ultramicroscopy 90~(2) (2002) 171--195.
\newblock \href {http://dx.doi.org/10.1016/S0304-3991(01)00134-6} {\path{doi:10.1016/S0304-3991(01)00134-6}}.
\newline\urlprefix\url{https://www.sciencedirect.com/science/article/pii/S0304399101001346}

\bibitem{martin_practical_2011}
A.~V. Martin, A.~I. Bishop, D.~M. Paganin, L.~J. Allen, \href{https://www.sciencedirect.com/science/article/pii/S0304399110002573}{Practical implementation of a direct method for coherent diffractive imaging}, Ultramicroscopy 111~(7) (2011) 777--781.
\newblock \href {http://dx.doi.org/10.1016/j.ultramic.2010.10.003} {\path{doi:10.1016/j.ultramic.2010.10.003}}.
\newline\urlprefix\url{https://www.sciencedirect.com/science/article/pii/S0304399110002573}

\bibitem{morgan_fast_2013}
A.~J. Morgan, A.~J. D’Alfonso, P.~Wang, H.~Sawada, A.~I. Kirkland, L.~J. Allen, \href{https://link.aps.org/doi/10.1103/PhysRevB.87.094115}{Fast deterministic single-exposure coherent diffractive imaging at sub-{\textbackslash}{AA}\{\}ngstr{\textbackslash}"om resolution}, Physical Review B 87~(9) (2013) 094115, publisher: American Physical Society.
\newblock \href {http://dx.doi.org/10.1103/PhysRevB.87.094115} {\path{doi:10.1103/PhysRevB.87.094115}}.
\newline\urlprefix\url{https://link.aps.org/doi/10.1103/PhysRevB.87.094115}

\bibitem{drenth_problem_1975}
A.~Drenth, A.~Huiser, H.~Ferwerda, \href{https://doi.org/10.1080/713819083}{The {Problem} of {Phase} {Retrieval} in {Light} and {Electron} {Microscopy} of {Strong} {Objects}}, Optica Acta: International Journal of Optics 22~(7) (1975) 615--628, publisher: Taylor \& Francis \_eprint: https://doi.org/10.1080/713819083.
\newblock \href {http://dx.doi.org/10.1080/713819083} {\path{doi:10.1080/713819083}}.
\newline\urlprefix\url{https://doi.org/10.1080/713819083}

\bibitem{songge_li_2025}
S.~Li, N.~Gauquelin, R.~H.~L. Lalandec, A.~Annys, C.~Gao, C.~Hofer, T.~J. Pennycook, J.~Verbeeck, \href{http://arxiv.org/abs/2505.09555}{Improving the low-dose performance of aberration correction in single sideband ptychography}\href {http://dx.doi.org/10.48550/arXiv.2505.09555} {\path{doi:10.48550/arXiv.2505.09555}}.
\newline\urlprefix\url{http://arxiv.org/abs/2505.09555}

\bibitem{yang_enhanced_2016}
H.~Yang, P.~Ercius, P.~D. Nellist, C.~Ophus, \href{https://www.sciencedirect.com/science/article/pii/S0304399116301966}{Enhanced phase contrast transfer using ptychography combined with a pre-specimen phase plate in a scanning transmission electron microscope}, Ultramicroscopy 171 (2016) 117--125.
\newblock \href {http://dx.doi.org/10.1016/j.ultramic.2016.09.002} {\path{doi:10.1016/j.ultramic.2016.09.002}}.
\newline\urlprefix\url{https://www.sciencedirect.com/science/article/pii/S0304399116301966}

\bibitem{hawkes_ptychography_2019}
J.~Rodenburg, A.~Maiden, \href{http://link.springer.com/10.1007/978-3-030-00069-1_17}{Ptychography}, in: P.~W. Hawkes, J.~C.~H. Spence (Eds.), Springer {Handbook} of {Microscopy}, Springer International Publishing, Cham, 2019, pp. 819--904, series Title: Springer Handbooks.
\newblock \href {http://dx.doi.org/10.1007/978-3-030-00069-1_17} {\path{doi:10.1007/978-3-030-00069-1_17}}.
\newline\urlprefix\url{http://link.springer.com/10.1007/978-3-030-00069-1_17}

\bibitem{maiden_superresolution_2011}
A.~M. Maiden, M.~J. Humphry, F.~Zhang, J.~M. Rodenburg, \href{https://opg.optica.org/abstract.cfm?URI=josaa-28-4-604}{Superresolution imaging via ptychography}, Journal of the Optical Society of America A 28~(4) (2011) 604.
\newblock \href {http://dx.doi.org/10.1364/JOSAA.28.000604} {\path{doi:10.1364/JOSAA.28.000604}}.
\newline\urlprefix\url{https://opg.optica.org/abstract.cfm?URI=josaa-28-4-604}

\bibitem{gerchberg_practical_1972}
R.~Gerchberg, W.~Saxton, A {Practical} {Algorithm} for the {Determination} of {Phase} from {Image} and {Diffraction} {Plane} {Pictures}, Optik 35.

\bibitem{gerchberg_super-resolution_1974}
R.~Gerchberg, \href{https://doi.org/10.1080/713818946}{Super-resolution through {Error} {Energy} {Reduction}}, Optica Acta: International Journal of Optics 21~(9) (1974) 709--720, publisher: Taylor \& Francis \_eprint: https://doi.org/10.1080/713818946.
\newblock \href {http://dx.doi.org/10.1080/713818946} {\path{doi:10.1080/713818946}}.
\newline\urlprefix\url{https://doi.org/10.1080/713818946}

\bibitem{melnyk_connections_2022}
O.~Melnyk, \href{https://doi.org/10.1007/s43670-022-00035-5}{On connections between {Amplitude} {Flow} and {Error} {Reduction} for phase retrieval and ptychography}, Sampling Theory, Signal Processing, and Data Analysis 20~(2) (2022) 16.
\newblock \href {http://dx.doi.org/10.1007/s43670-022-00035-5} {\path{doi:10.1007/s43670-022-00035-5}}.
\newline\urlprefix\url{https://doi.org/10.1007/s43670-022-00035-5}

\bibitem{guizar-sicairos_phase_2008}
M.~Guizar-Sicairos, J.~R. Fienup, \href{https://opg.optica.org/oe/abstract.cfm?uri=oe-16-10-7264}{Phase retrieval with transverse translation diversity: a nonlinear optimization approach}, Optics Express 16~(10) (2008) 7264--7278, publisher: Optica Publishing Group.
\newblock \href {http://dx.doi.org/10.1364/OE.16.007264} {\path{doi:10.1364/OE.16.007264}}.
\newline\urlprefix\url{https://opg.optica.org/oe/abstract.cfm?uri=oe-16-10-7264}

\bibitem{odstrcil_iterative_2018}
M.~Odstrčil, A.~Menzel, M.~Guizar-Sicairos, \href{https://opg.optica.org/oe/abstract.cfm?uri=oe-26-3-3108}{Iterative least-squares solver for generalized maximum-likelihood ptychography}, Optics Express 26~(3) (2018) 3108--3123, publisher: Optica Publishing Group.
\newblock \href {http://dx.doi.org/10.1364/OE.26.003108} {\path{doi:10.1364/OE.26.003108}}.
\newline\urlprefix\url{https://opg.optica.org/oe/abstract.cfm?uri=oe-26-3-3108}

\bibitem{thibault_maximum-likelihood_2012}
P.~Thibault, M.~Guizar-Sicairos, \href{https://dx.doi.org/10.1088/1367-2630/14/6/063004}{Maximum-likelihood refinement for coherent diffractive imaging}, New Journal of Physics 14~(6) (2012) 063004, publisher: IOP Publishing.
\newblock \href {http://dx.doi.org/10.1088/1367-2630/14/6/063004} {\path{doi:10.1088/1367-2630/14/6/063004}}.
\newline\urlprefix\url{https://dx.doi.org/10.1088/1367-2630/14/6/063004}

\bibitem{madsen_abtem_2021}
J.~Madsen, T.~Susi, \href{https://open-research-europe.ec.europa.eu/articles/1-24/v1}{The {abTEM} code: transmission electron microscopy from first principles}, Open Research Europe 1 (2021) 24.
\newblock \href {http://dx.doi.org/10.12688/openreseurope.13015.1} {\path{doi:10.12688/openreseurope.13015.1}}.
\newline\urlprefix\url{https://open-research-europe.ec.europa.eu/articles/1-24/v1}

\bibitem{huang_solar-driven_2020}
H.~Huang, B.~Pradhan, J.~Hofkens, M.~B.~J. Roeffaers, J.~A. Steele, \href{https://doi.org/10.1021/acsenergylett.0c00058}{Solar-{Driven} {Metal} {Halide} {Perovskite} {Photocatalysis}: {Design}, {Stability}, and {Performance}}, ACS Energy Letters 5~(4) (2020) 1107--1123, publisher: American Chemical Society.
\newblock \href {http://dx.doi.org/10.1021/acsenergylett.0c00058} {\path{doi:10.1021/acsenergylett.0c00058}}.
\newline\urlprefix\url{https://doi.org/10.1021/acsenergylett.0c00058}

\bibitem{dey_state_2021}
A.~Dey, J.~Ye, A.~De, E.~Debroye, S.~K. Ha, E.~Bladt, A.~S. Kshirsagar, Z.~Wang, J.~Yin, Y.~Wang, L.~N. Quan, F.~Yan, M.~Gao, X.~Li, J.~Shamsi, T.~Debnath, M.~Cao, M.~A. Scheel, S.~Kumar, J.~A. Steele, M.~Gerhard, L.~Chouhan, K.~Xu, X.-g. Wu, Y.~Li, Y.~Zhang, A.~Dutta, C.~Han, I.~Vincon, A.~L. Rogach, A.~Nag, A.~Samanta, B.~A. Korgel, C.-J. Shih, D.~R. Gamelin, D.~H. Son, H.~Zeng, H.~Zhong, H.~Sun, H.~V. Demir, I.~G. Scheblykin, I.~Mora-Seró, J.~K. Stolarczyk, J.~Z. Zhang, J.~Feldmann, J.~Hofkens, J.~M. Luther, J.~Pérez-Prieto, L.~Li, L.~Manna, M.~I. Bodnarchuk, M.~V. Kovalenko, M.~B.~J. Roeffaers, N.~Pradhan, O.~F. Mohammed, O.~M. Bakr, P.~Yang, P.~Müller-Buschbaum, P.~V. Kamat, Q.~Bao, Q.~Zhang, R.~Krahne, R.~E. Galian, S.~D. Stranks, S.~Bals, V.~Biju, W.~A. Tisdale, Y.~Yan, R.~L.~Z. Hoye, L.~Polavarapu, \href{https://doi.org/10.1021/acsnano.0c08903}{State of the {Art} and {Prospects} for {Halide} {Perovskite} {Nanocrystals}}, ACS Nano 15~(7) (2021) 10775--10981, publisher: American Chemical Society.
\newblock \href {http://dx.doi.org/10.1021/acsnano.0c08903} {\path{doi:10.1021/acsnano.0c08903}}.
\newline\urlprefix\url{https://doi.org/10.1021/acsnano.0c08903}

\bibitem{bunk_influence_2008}
O.~Bunk, M.~Dierolf, S.~Kynde, I.~Johnson, O.~Marti, F.~Pfeiffer, \href{https://www.sciencedirect.com/science/article/pii/S0304399107001969}{Influence of the overlap parameter on the convergence of the ptychographical iterative engine}, Ultramicroscopy 108~(5) (2008) 481--487.
\newblock \href {http://dx.doi.org/10.1016/j.ultramic.2007.08.003} {\path{doi:10.1016/j.ultramic.2007.08.003}}.
\newline\urlprefix\url{https://www.sciencedirect.com/science/article/pii/S0304399107001969}

\bibitem{cowley_scattering_1957}
J.~M. Cowley, A.~F. Moodie, \href{https://journals.iucr.org/q/issues/1957/10/00/a02113/}{The scattering of electrons by atoms and crystals. {I}. {A} new theoretical approach}, Acta Crystallographica 10~(10) (1957) 609--619, publisher: International Union of Crystallography.
\newblock \href {http://dx.doi.org/10.1107/S0365110X57002194} {\path{doi:10.1107/S0365110X57002194}}.
\newline\urlprefix\url{https://journals.iucr.org/q/issues/1957/10/00/a02113/}

\bibitem{goodman_numerical_1974}
P.~Goodman, A.~F. Moodie, \href{https://journals.iucr.org/a/issues/1974/02/00/a10626/}{Numerical evaluations of {N}-beam wave functions in electron scattering by the multi-slice method}, Acta Crystallographica Section A: Crystal Physics, Diffraction, Theoretical and General Crystallography 30~(2) (1974) 280--290, publisher: International Union of Crystallography.
\newblock \href {http://dx.doi.org/10.1107/S056773947400057X} {\path{doi:10.1107/S056773947400057X}}.
\newline\urlprefix\url{https://journals.iucr.org/a/issues/1974/02/00/a10626/}

\bibitem{ishizuka_new_1977}
K.~Ishizuka, N.~Uyeda, \href{https://journals.iucr.org/a/issues/1977/05/00/a14220/}{A new theoretical and practical approach to the multislice method}, Acta Crystallographica Section A: Crystal Physics, Diffraction, Theoretical and General Crystallography 33~(5) (1977) 740--749, publisher: International Union of Crystallography.
\newblock \href {http://dx.doi.org/10.1107/S0567739477001879} {\path{doi:10.1107/S0567739477001879}}.
\newline\urlprefix\url{https://journals.iucr.org/a/issues/1977/05/00/a14220/}

\bibitem{andrew_maiden_ptychography_2025}
A.~Andrew~Maiden, \href{https://github.com/andyMaiden/SheffieldPtycho}{Ptychography algorithms from {Sheffield} {University}}, original-date: 2023-12-22T11:31:17Z (May 2025).
\newline\urlprefix\url{https://github.com/andyMaiden/SheffieldPtycho}

\bibitem{noauthor_pyptychostem_2024}
\href{https://gitlab.com/pyptychostem/pyptychostem}{{PyPtychoSTEM} / {pyPtychoSTEM} · {GitLab}} (May 2024).
\newline\urlprefix\url{https://gitlab.com/pyptychostem/pyptychostem}

\bibitem{hofer_phase_2024}
C.~Hofer, C.~Gao, T.~Chennit, B.~Yuan, T.~J. Pennycook, \href{https://www.sciencedirect.com/science/article/pii/S0304399124000019}{Phase offset method of ptychographic contrast reversal correction}, Ultramicroscopy 258 (2024) 113922.
\newblock \href {http://dx.doi.org/10.1016/j.ultramic.2024.113922} {\path{doi:10.1016/j.ultramic.2024.113922}}.
\newline\urlprefix\url{https://www.sciencedirect.com/science/article/pii/S0304399124000019}

\end{thebibliography}

\end{document}